\documentclass[12pt,twocolumn]{aastex6}
\usepackage{amssymb,amsmath}
\usepackage{hyperref}
\usepackage{ulem}
\usepackage{color}
\usepackage{natbib}
\usepackage{fancyhdr}

\slugcomment{}

\shorttitle{Size evolution of quiescent F2-HSC galaxies}
\shortauthors{Damjanov et al.}

\begin{document}

\title{Quiescent Galaxy Size and Spectroscopic Evolution: Combining HSC Imaging and Hectospec Spectroscopy}

\author{Ivana Damjanov\altaffilmark{1,2,3}, H. Jabran Zahid\altaffilmark{3}, Margaret J. Geller\altaffilmark{3}, Yousuke Utsumi\altaffilmark{4,5}, Jubee Sohn\altaffilmark{3}, Harrison Souchereau\altaffilmark{1}}
\altaffiltext{1}{Department of Astronomy and Physics, Saint Mary's University, 923 Robie Street, Halifax, NS B3H 3C3, Canada; \href{mailto:Ivana.Damjanov@smu.ca}{Ivana.Damjanov@smu.ca}}
\altaffiltext{2}{Canada Research Chair}
\altaffiltext{3}{Harvard-Smithsonian Center for Astrophysics, 60 Garden Street, Cambridge, MA 02138, USA; \href{mailto:idamjanov@cfa.harvard.edu}{idamjanov@cfa.harvard.edu}}
\altaffiltext{3}{Kavli Institute for Particle Astrophysics and Cosmology, SLAC National Accelerator Laboratory, Stanford University, Stanford, CA 94305, USA}
\altaffiltext{4}{Hiroshima Astrophysical Science Center, Hiroshima University, 1-3-1 Kagamiyama, Higashi-Hiroshima, Hiroshima, 739-8526, Japan}

\begin{abstract}
We explore the relations between size, stellar mass and average stellar population age (indicated by D$_n4000$ indices) for a sample of $\sim11000$ intermediate-redshift galaxies from the SHELS spectroscopic survey \citep{Geller2014} augmented by high-resolution Subaru Telescope Hyper Suprime-Cam imaging. In the redshift interval $0.1<z<0.6$, star forming galaxies are on average larger than their quiescent counterparts. The mass-complete sample of $\sim3500$ $M_\ast>10^{10}\, M_\sun$ quiescent galaxies shows that the average size of a $10^{11}\, M_\sun$ quiescent galaxy increases by $\lesssim25\%$ from $z\sim0.6$ to $z\sim0.1$. This growth rate is a function of stellar mass:  the most massive ($M_\ast>10^{11}\, M_\sun$) galaxies grow significntly more slowly in size than an order of magnitude less massive quiescent systems that grow by 70\% in the $0.1\lesssim z\lesssim0.3$ redshift interval. For $M_\ast<10^{11}\, M_\sun$ galaxies  age and size are anti-correlated at fixed mass; more massive quiescent systems show no significant trend in size with average stellar population age. The evolution in absolute and fractional abundances of quiescent systems at intermediate redshift are also a function of galaxy stellar mass. The suite of evolutionary trends suggests that galaxies more massive than $\sim10^{11}\, M_\sun$ have mostly assembled their mass by $z\sim0.6$.  Quiescent galaxies with lower stellar masses show more complex evolution that is characterized by a combination of individual quiescent galaxy size growth (through mergers) and an increase in the size of newly quenched galaxies joining the population at later times (progenitor bias).  The low-mass population ($M_\ast\sim10^{10}\, M_\sun$) grows predominantly as a result of progenitor bias.  For more massive ($M_\ast\sim5\times10^{10}\, M_\sun$) quiescent galaxies, (predominantly minor) mergers and progenitor bias make more comparable contributions to the size growth. At intermediate redshift quiescent size growth is mass-dependent; the most massive ($M_\ast>10^{11}\, M_\sun$) galaxies experience the least rapid increase in size from $z\sim0.6$ to $z\sim0.1$. 

\end{abstract}


\section{Introduction}

Galaxy size is a fundamental galaxy property that quantifies the  distribution of its stellar mass. In combination with the concentration of surface brightness profiles and stellar velocity dispersion, galaxy sizes provide a connection between the luminous content and dynamical mass of quiescent galaxies \citep[e.g., ][]{Zahid2017}. The distribution and redshift evolution of sizes for quiescent  systems  offer clues about the history of their mass assembly.

The size or half-light (or effective) radius of a quiescent galaxy is directly related to its surface brightness \citep{Kormendy1977}. This relation is one projection of the Fundamental Plane, a tight correlation between luminosity, velocity dispersion, and size  \citep{Djorgovski1987,Dressler1987}. Using galaxy stellar mass rather than surface brightness accounts for mass-to-light ratio variations and enables study of the relation over a broad redshift range. 

A large suite of observational studies  map the size -- stellar mass relation for quiescent galaxies and its evolution over the $0\lesssim z\lesssim3$ redshift interval \citep[e.g.,][]{Shen2003,Trujillo2004,Zirm2007,Toft2007,Trujillo2007,Buitrago2008,Guo2009,vanDokkum2010,Damjanov2011,Newman2012,Cimatti2012,HuertasCompany2013,vanderWel2014,Lange2015,Sweet2017}. A majority of these studies are based on  \textit{Hubble Space Telescope} (HST) imaging.  Quiescent systems at $z\sim1.5$ are on average a factor of $2-3$ smaller than their counterparts of the same mass at $z\sim0$ \citep{Daddi2005}. Furthermore, these studies revealed a population of extremely massive compact quiescent galaxies at high redshift. These compact systems  are $5-10$~times smaller than similarly massive local quiescent galaxies \citep[e.g,][]{vanDokkum2008}. 

Direct comparisons of equivalently selected compact samples at high redshift ($z>1$) and in the local volume (based on Sloan Digital Sky Survey (SDSS) imaging) imply a drastic decline in the abundance of massive compact quiescent systems \citep[e.g.][]{Trujillo2009,Cassata2011,Cassata2013,vanderWel2014}. However, other $z\sim0$ studies of massive quiescent compacts report number densities  similar to the abundances of $z>1$ systems \citep{Valentinuzzi2010a, Poggianti2013a,Poggianti2013b}. 

Recent  investigations of quiescent samples at $0.2<z<0.8$ link 
the local and $z > 1$ samples \citep{Carollo2013,Damjanov2015a,Tortora2016,Charbonnier2017}. These studies confirm that compact systems experience at most a moderate change in number density from $z\lesssim1$ to $z\sim0$. Although some compact systems may grow, new massive quiescent compact galaxies may also form at $z<1$ \citep{Zahid2016b}. Dense spectroscopic surveys \citep[e.g.,][]{Damjanov2018} at intermediate-redshift  show that massive compact galaxies are the extension of a continuum of structural, stellar population, dynamical, and environmental properties that characterize the full  quiescent population \citep{Damjanov2015b,Zahid2016a}. In combination with their large stellar mass, mild number density evolution of massive compacts suggests that processes driving the evolution in size of individual systems and/or the growth in the average size of quiescent population may depend on galaxy stellar mass.

Evolutionary trends in 1) the parameters that define the quiescent galaxy size -- stellar mass relation, and 2) the number density of quiescent systems  provide  constraints on models of galaxy mass assembly. Theoretical models of individual galaxy size growth include the effects of 1) major mergers between gas-poor galaxies of similar stellar mass, 2) minor merger or accretion of low surface brightness objects, and 3) adiabatic expansion after significant mass loss \citep[e.g.,][]{Hopkins2010a}.  Comparison between the observed properties of $z\gtrsim1$ and local quiescent systems suggest that minor mergers dominate size growth of quiescent galaxies \citep{White2007,Bezanson2009,Newman2012,McLure2013,vandeSande2013}. For stellar masses $M_\ast<10^{11}\, M_\sun$ the growth is further altered by the addition of larger newly quenched galaxies at $z < 1$ \citep{Carollo2013,Fagioli2016}. The most massive galaxies ($M_\ast>10^{11}\, M_\sun$) become quiescent at earlier epochs \citep[downsizing, ][]{Cowie1996}.

The stellar population age \citep[e.g.,][]{Belli2015,Fagioli2016} provides an additional powerful constraint on the evolutionary processes affecting the quiescent population.  The age of the stellar population correlates with the structural properties and stellar masses of quiescent galaxies at $z\sim0$: at fixed stellar mass older quiescent systems are smaller \citep{Zahid2017}. The spectral indicator D$_n4000$ \citep[e.g,][]{Kauffmann2003} provides a convenient measure of the quiescent population age. Complete redshift surveys that include D$_n4000$ thus enable selection and investigation of systems that become quiescent at the high redshift limit of the survey and evolve over the redshift range of the survey.

Analysis of the relations between quiescent galaxy size growth and the processes that drive it requires a combination of large-area high-quality imaging and dense spectroscopy. Here we   trace  quiescent galaxy size evolution for redshifts  $0.1\leqslant z\leqslant0.6$  as a function of galaxy stellar mass and average stellar population age measured by the D$_n4000$ index). In Section~\ref{data} we review the intermediate-redshift spectroscopic survey \citep[SHELS,][]{Geller2014} and the associated HSC high-resolution imaging \citep{Utsumi2016}. Section~\ref{sizemeas} describes  the measurement of galaxy sizes and the distribution of  sizes  as a function of redshift and stellar mass. We then explore the relations among average age, stellar mass, and size of quiescent systems  (Section~\ref{globquiescent}) and extract evolutionary constraints on the size growth of quiescent galaxies (Section~\ref{ec}). We discuss the results in Section~\ref{dis} and conclude in Section~\ref{conc}.  We adopt the standard cosmology $(H_0,\, \Omega_m,\, \Omega_ \Lambda) = (70$~km~s$^{-1}$~Mpc$^{-1},\, 0.3,\, 0.7)$ and AB magnitudes throughout. We use the \citet{Chabrier2003} initial mass function (IMF) in computing stellar masses.

\section{Dataset}\label{data}

\subsection{HSC Imaging}\label{imaginig}

We measure galaxy sizes using Subaru Hyper Suprime-Cam \citep[HSC;][]{Miyazaki2012} $i$-band images of the $2\times2~{\rm deg}^2$ region covering the Deep Lens Survey (DLS) \citep{Wittman2002} field F2: $(\alpha_c, \delta_c)=(9^{h}18^{m}0^{s},  +30^{\circ}00'00")$. \citet{Utsumi2016} describe the HSC observing procedure and image processing in detail; we provide a short summary here.

The F2 HSC image includes 18 pointings each with a 240 second exposure. The pointings overlap and extend beyond the 4~deg$^2$ footprint of the F2 field to yield uniform depth \citep[see~Figure~1 in][]{Utsumi2016}. Galaxy number counts show that the HSC F2 image provides a complete catalog of extended sources to a limiting $i-$band magnitude of $i\sim25$. The typical seeing full with at half maximum (FWHM) is $\sim0\farcs6$.   

The {\it hscPipe} system \citep{Bosch2018} is the standard pipeline, developed for the HSC Subaru Strategic Program (SSP), that performs reduction of individual chips, mosaicking, and image stacking. We use the SExtractor software \citep{Bertin1996} to measure photometric parameters, that include galaxy half-light radius, ellipticity, and S\'{e}rsic index, from stacked HSC images. These parameters are based
on two-dimensional (2D) modelling of the galaxy surface brightness profile following a three-step process: a) in its first run SExtractor provides a catalog of sources that includes star-galaxy separation; b) the PSFex software \citep{Bertin2011} combines point sources from the initial SExtractor catalog to construct a set of spatially varying Point Spread Functions (PSFs) that are used as input parameters for c) the second SExtractor run to provide a catalog with morphological parameters for all detected sources. 

Here we employ single S\'{e}rsic profile models \citep{Sersic1968} to estimate the  sizes of the SHELS F2 galaxies. This approach enables the use of existing catalogs (based on lower-resolution and less sensitive imaging) for input parameter values and for estimation of external errors in the parameters we obtain with SExtractor (Section~\ref{sizemeas}).

\subsection{SHELS spectroscopy}\label{spec}

DLS imaging provided the photometric catalog for the complete redshift survey of the F2 field \citep[SHELS,][]{Geller2014} carried out with the Hectospec wide-field multi-object spectrograph mounted on the MMT. The redshift survey is 95\% complete to a limiting magnitude $R = 20.6$ and it covers 3.98 deg$^2$. 

\citet{Geller2014} provide detailed description of the SHELS F2 spectroscopy. Here we briefly quantify the galaxy sample and describe the spectro-photometric parameters from \citet{Geller2014} that we use in our analysis.

The complete SHELS F2 sample includes 13327 galaxies with $R_\mathrm{mag}\leq20.6$. The $D_n$4000 index is the ratio of flux (in $f_\nu$ units) in the $4000-4100 {\rm \AA}$ and $3850-3950{\rm \AA}$ bands \citep{Balogh1999}. The median signal-to-noise ratio ($S/N$) of SHELS F2 spectra around 4000\, \AA\ is $\sim6$ per resolution element and 90\% of galaxies have $\sim3<S/N<15$ per resolution element. \citet{Geller2014} report the $D_n$4000 index and stellar mass measurements for 10730 galaxies in this sample ($\gtrsim80\%$). For the SHELS data, the typical fractional error
in the $D_n$4000 (based on 1468 repeat measurements) is 4.5\%.

The strength of the $4000$~\AA\ break is smaller for systems dominated by young stellar populations and it increases with the stellar population age. Large spectroscopic surveys demonstrate that the D$_n4000$ index distribution is strongly bimodal, with a clear division between quiescent and star-forming galaxies at D$_n4000\sim1.5$ \citep{Kauffmann2003, Vergani2008, Woods2010, Geller2014, Damjanov2018}. Following \citet{Woods2010} we use D$_n4000 = 1.5$ as the dividing line between the two populations.

We check whether the  results could be biased by the D$_n4000$ selection. We emphasize that 
D$_n4000$ is a powerful evolutionary marker because it is insensitive to reddening and, in contrast with galaxy colors, does not require a $K-$correction. Quiescent galaxy selection based on the spectral index cut (D$_n4000>1.5$) agrees well with  rest-frame $UVJ$ color selection based on fitting of SEDs obtained from 30 photometric bands in the $0.15-24\, \mu$m wavelength range \citep[hCOSMOS, ][]{Damjanov2018}. Furthermore, hCOSMOS galaxies with D$_n4000>1.5$ follow the tight relation between size, velocity dispersion, and stellar mass surface density \citep[so-called Fundamental Plane,][]{Zahid2016a}. In the parameter space defined by model-independent diagnostics of galaxy morphology (asymmetry, concentration, Gini coefficient, and second-order moment of the brightest 20\% of galaxy pixels) the majority of  $1.5<\mathrm{D}_n4000<1.6$ hCOSMOS galaxies ($\sim70\%$) occupy the same regions as elliptical and bulge-dominated galaxies classified by the Zurich Estimator of Structural Types \citep[ZEST, ][]{Scarlata2007}. For D$_n4000>1.6$ hCOSMOS galaxies the overlap with ZEST-selected elliptical and bulge-dominated galaxies is $\sim90\%$. When we repeat the analysis of F2-HSC galaxy sample using the {f16}D$_n4000>1.6$ quiescent selection our results are robust to this upward shift in the spectral index cut.

Stellar masses of SHELS F2 galaxies are based on  SDSS five band photometry. \citet{Geller2014} fit the observed photometry with \texttt{Le Phare}\footnote{\url{http://www.cfht.hawaii.edu/$\sim$arnouts/LEPHARE/cfht\_lephare/ lephare.html}} \citep{Arnouts1999, Ilbert2006} using the stellar population synthesis models of \citet*{Bruzual2003} with \citet{Chabrier2003} IMF, two metallicites (0.4 and 1 solar), a set of exponentially decreasing star formation rates, and \citet{Calzetti2000} extinction law. The best-fit models provide the mass-to-light ratio, a scaling factor that transforms the observed luminosity into the galaxy stellar mass (i.e., its current living stellar mass).

\section{HSC sizes}\label{sizemeas}

\subsection{Method}

\begin{figure}
\begin{centering}
%\hspace*{-0.35in}
\includegraphics[scale=0.2]{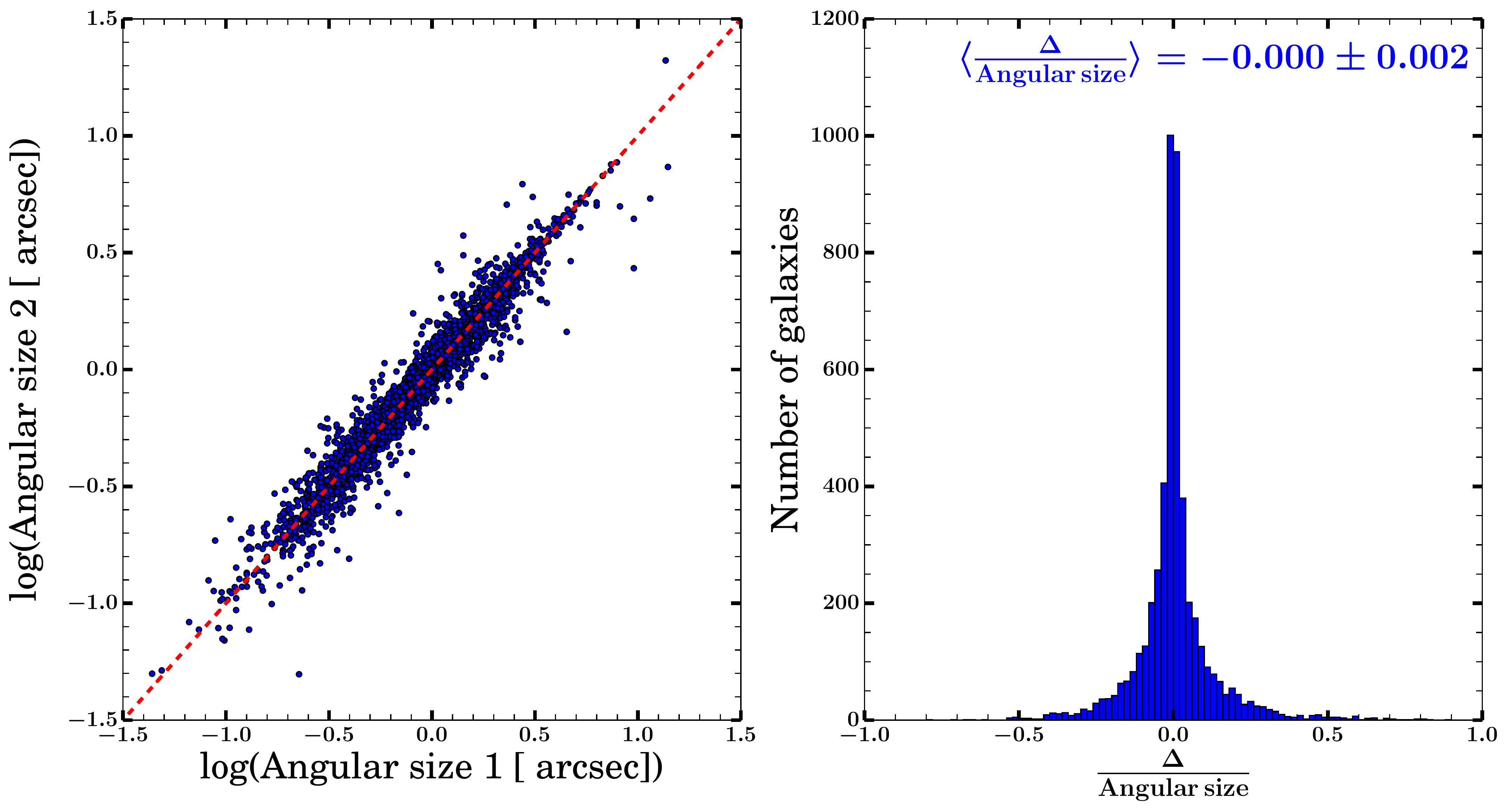}
\includegraphics[scale=0.2]{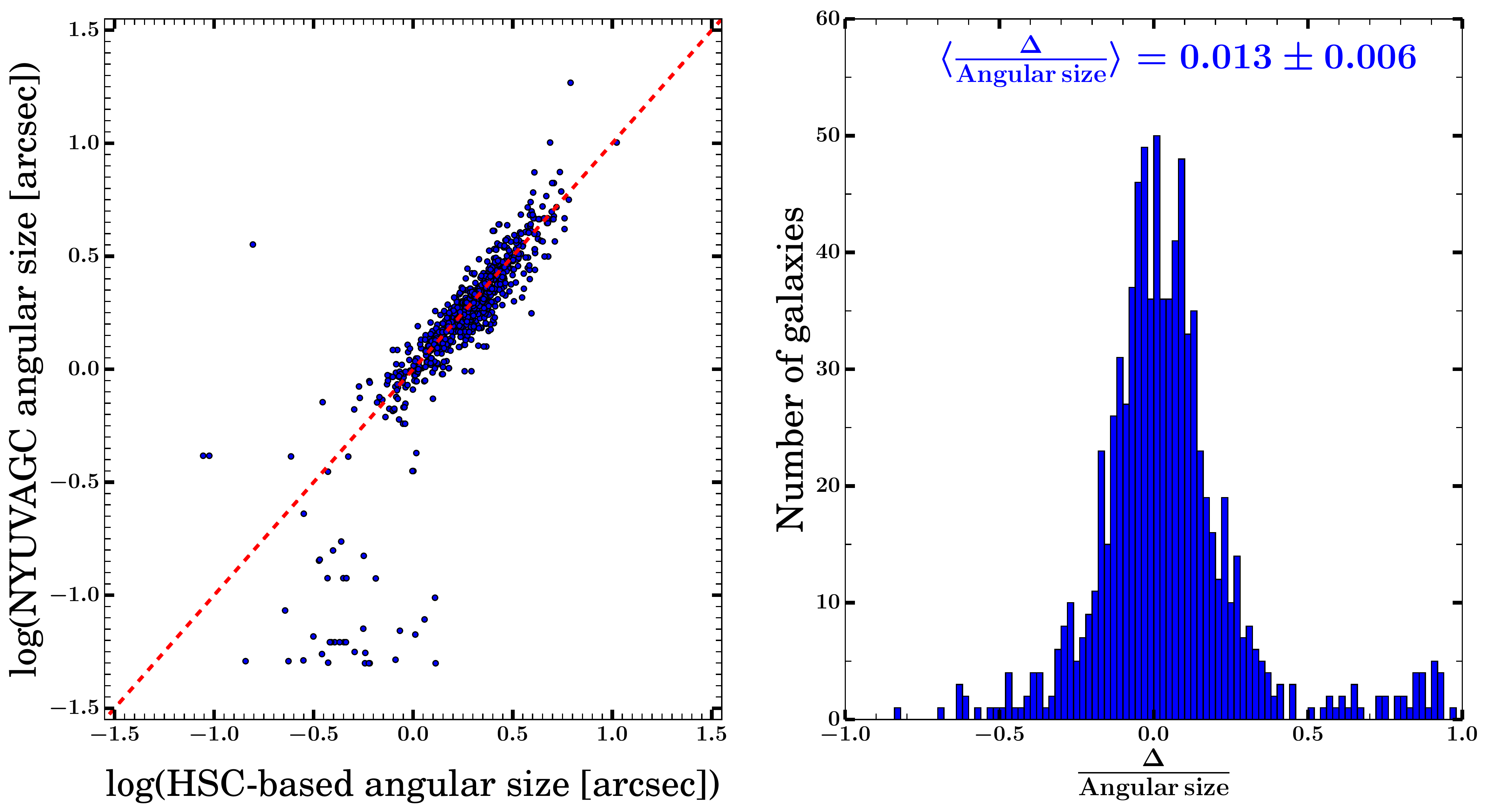}
\caption{Upper panels: Comparison between 4999 repeat angular size measurements ($r_e\times\sqrt{b/a}$) for F2-HSC galaxies. Lower panels: Comparison between HSC  and NYUVAG angular sizes for 796  SHELS F2 galaxies. The left-hand panels compare individual measurements. The right-hand panels show histograms of relative differences. Values on $x-$axis correspond to the differences between two size measurements ($\Delta$) as a fraction of the size listed in Table~\ref{table1}; in the lower right-hand panel $\Delta=\mathrm{Size_{HSC}}-\mathrm{Size_{NYU}}$. The legend indicates the mean relative offset and the error in the offset. \label{f1}}  
\end{centering}
\end{figure} 

\begin{figure*}
\begin{centering}
\includegraphics[scale=0.35]{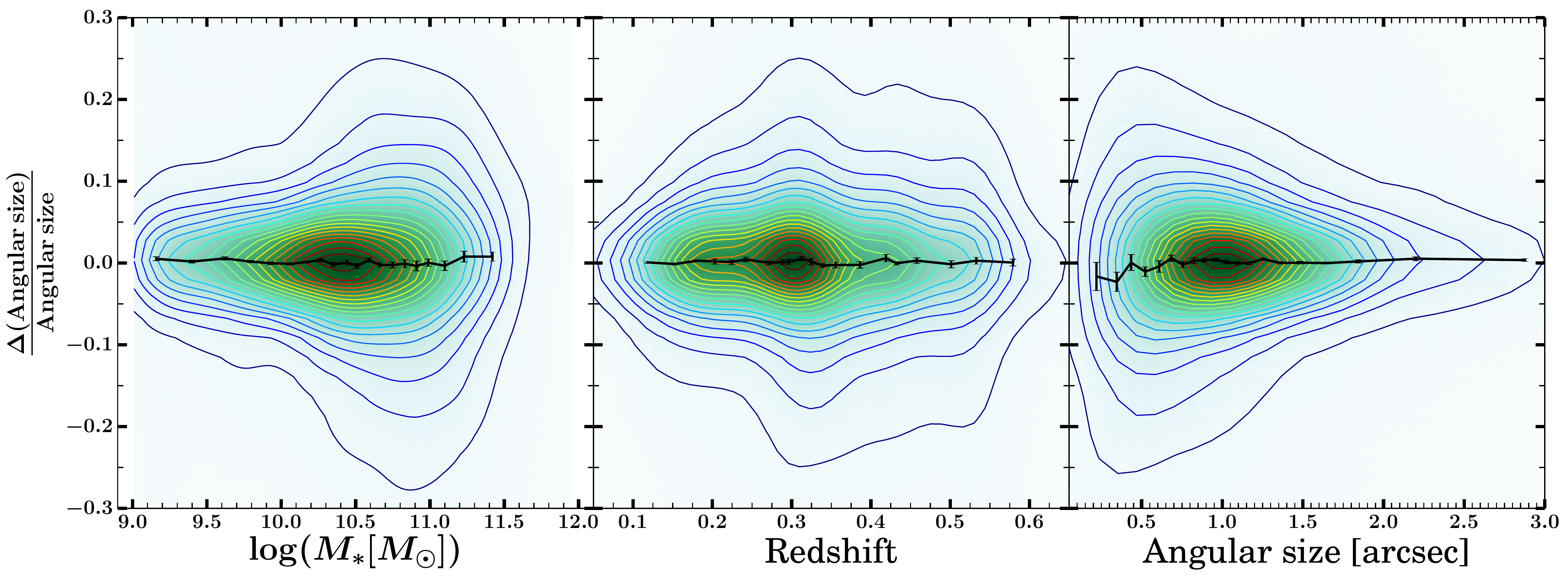}
\caption{Relative difference between multiple size measurements for 4999 F2-HSC galaxies as a function of galaxy stellar mass (left panel), redshift (central panel), and measured size (right panel). Green color maps in each panel show 2D density histograms that we smooth using a Gaussian kernel density estimator with bandwidth determined by the standard Scott's rule \citep{Scott2015}. Colored contours correspond to $5-95\%$ of the maximum density. Black lines connect median relative size offsets in equally populated bins of galaxy stellar mass, redshift, and size. Errors are bootstrapped.\label{f2}}    
\end{centering}
\end{figure*} 

\begin{figure}
\begin{centering}
%\hspace*{-0.35in}
\includegraphics[scale=0.35]{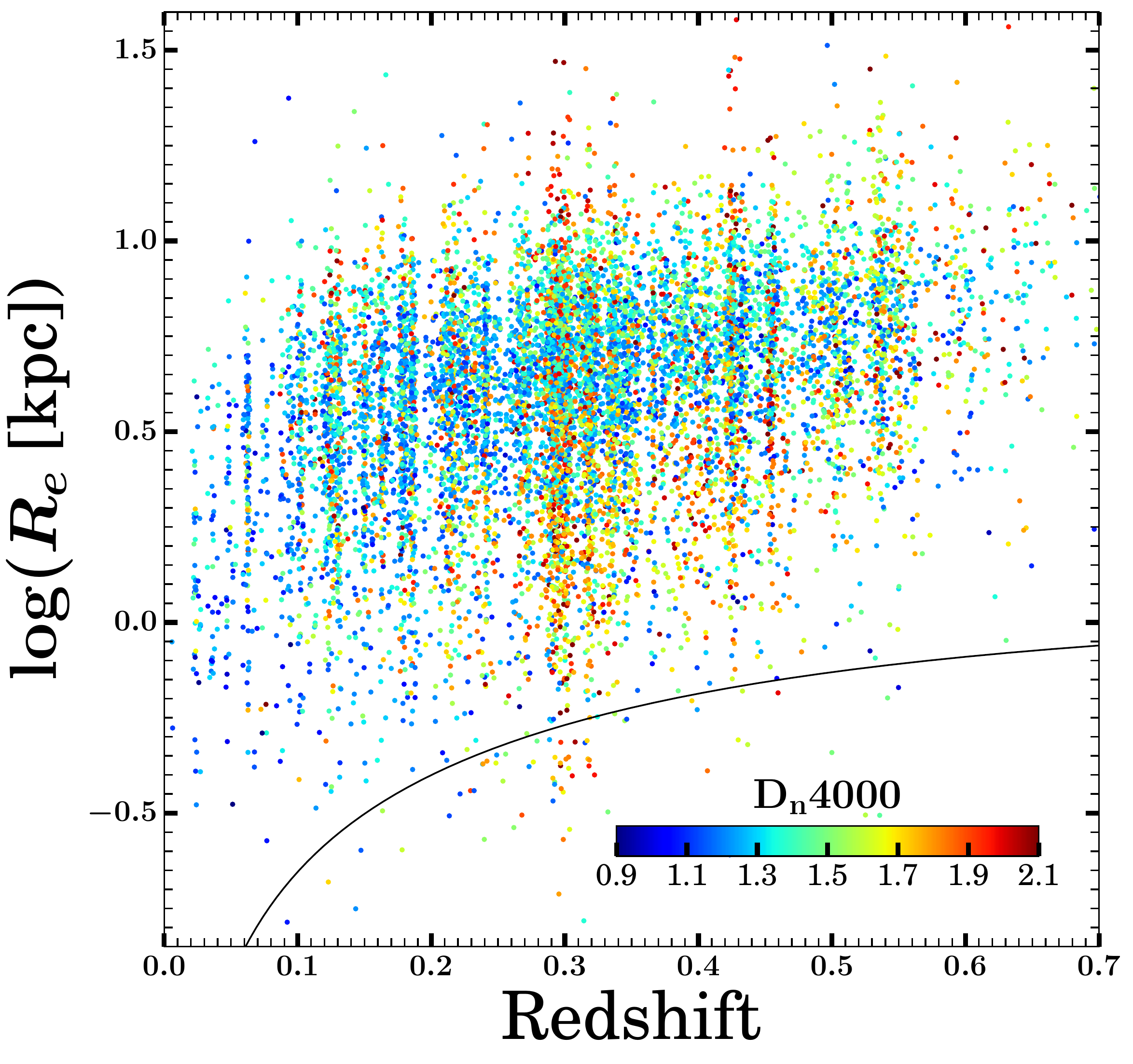}
\caption{Major axis radius as a function of redshift. Points are color-coded  by D$_n4000$. The black solid line shows the redshift-dependent limit on the major axis radius measurement set by the seeing. The points below the line correspond to galaxy sizes that are smaller than the seeing-defined limit \label{f3}}  
\end{centering}
\end{figure}

We obtain morphological parameters  by fitting the radial dependence of the galaxy surface brightness with the intensity model in the HSC $i-$ band:

\begin{equation}\label{eq:sersic}
I(r)=I_0 \exp\left\{-b_n\left(\frac{r}{r_e}\right)^{1/n}\right\}
\end{equation}

\noindent \citep{Sersic1968}. The free parameters of this model are: the central intensity (surface brightness) $I_0$,  the half-light radius along the galaxy major axis $r_e$ (in arcseconds), and the S\'ersic index $n$ that describes the concentration of the model profile. The coefficient $b_n$, a function of $n$, is defined to ensure that $r_e$ encloses the half of the total galaxy light.

The surface brightness profile fitting routine is part of the SExtractor software\footnote{\url{http://sextractor.readthedocs.io/en/latest/Model.html}} \citep[Section~\ref{imaginig}][]{Bertin1996}. SExtractor fits radial profiles of extended sources in the HSC image with a surface brightness model (based on S\'ersic profile, Eq.~\ref{eq:sersic}) convolved with the local PSF provided by the PSFEx software\footnote{\url{http://psfex.readthedocs.io/en/latest/Working.html}} \citep{Bertin2011}. The PSF model at the position of an extended source is a linear combination of basic vectors that best fits the observed point source profiles in the vicinity of that position. 

Within the \texttt{SPHEROID} model, the SExtractor fitting procedure provides a list of measurements including the S\'ersic profile parameters (Eq.~\ref{eq:sersic}) and the best-fit model axial ratio $b/a$ (SExtractor model parameter \texttt{SPHEROID\_ASPECT\_WORLD}). We derive the angular diameter distance at the spectroscopic redshift of each galaxy to translate the galaxy angular size $r_e$ (SExtractor model parameter \texttt{SPHEROID\_SCALE\_WORLD}) into the  major axis radius $R_e$ in kpc \citep[e.g.,][and references therein]{Hogg1999}. To comply with the size measurements  in the literature, we further combine the half-light radius $R_e$ with the axial ratio to derive the circularized half-light radius $R_{e,c}=R_e\times\sqrt{b/a}$.

We obtain structural parameters for 12823 SHELS F2 galaxies with $R_\mathrm{mag}\leq20.6$. To combine the HSC photometry with the spectroscopy we require that the separation between the best-matched photometric and spectroscopic target positions are less than the HSC imaging resolution ($0\farcs6$). Table~\ref{table1} lists the half-light radii $r_e$ and axial ratios $b/a$ for all 12823 galaxies with measured sizes in the magnitude limited SHELS F2 sample (10515 galaxies also have stellar mass and D$_n4000$ index measurements).  

A number of SHELS F2 sources have more than one size measurement within $0\farcs6$ because they are in regions where multiple HSC pointings overlap \citep[as shown in Figure~1 of ][]{Utsumi2016}. These repeat measurements provide a unique opportunity to estimate the size measurement error independent of  the statistical error in the individual fits. We compare these 4999 repeat measurements (upper panels of Figure~\ref{f1}) and calculate the typical relative internal error, $\sigma_{int}=5\%$,  by requiring that 68\% of the measurement differences are within $\sqrt{2}\sigma_{int}$. 

We  also compare two independent size measurements for a subsample of SHELS F2 galaxies (lower panels of Figure~\ref{f1}). The NYU Value-added Galaxy Catalog \citep[NYUVAGC,][]{Blanton2005, Padmanabhan2008} provides SDSS-based size measurements in $ugriz$ bands for 796 SHELS F2 galaxies. These measurements  are based on the S\'ersic profile fits to the observed azimuthally averaged galaxy radial profiles\footnote{NYUVAGC $i-$band size measurements are consistent with HST F814W-based measurements for galaxies in the COSMOS field \citep{Zahid2016a}.}. We directly compare $i$~band-based NYUVAGC sizes\footnote{reported in \texttt{SERSIC\_R50} column of the catalog} with the HSC measurements of circularized angular radii. Except for a small number of outliers with sizes far below the limiting resolution of SDSS imaging ($\log(\mathrm{ NYUVAGC\, angular\, size})\lesssim-1$), the two size estimates agree remarkably well. The average relative size difference is marginally consistent (within $\sim2\sigma$) with no offset between the two measurements (lower right panel of Figure~\ref{f1}). Assuming  an equal division of errors between SDSS- and HSC-based size measurements, a requirement that 68\% of the measurement differences are within $\sqrt{2}\sigma_{ext}$ provides a typical relative external error estimate of $10\%$.

Regions of the F2 field that are imaged multiple times in HSC enable us uniquely not only to estimate internal errors on size measurements, but also to probe some of the biases that might affect these measurements \citep[e.g.,][and references therein]{Faisst2017}. We use F2-HSC sources with multiple size measurements to test whether the estimated internal error shows any trends with galaxy properties. Relative size differences between multiple measurements are confined to $\pm25\%$ for the full range of galaxy stellar masses, redshifts, and sizes (colored contours in three panels of Figure~\ref{f2}). Even more important for this analysis, the median size offset (black solid lines with error bars in Figure~\ref{f2}) is consistent with zero (within $1-2\, \sigma$ uncertainty) in all equally populated bins for all three galaxy parameters. Thus we conclude that the F2-HSC size measurements do not exhibit any systematics with galaxy stellar mass, redshift, or size. We emphasize that only the relative sizes (i.e., change is size) are  important in the size growth analysis; Figure~\ref{f2} shows that the relative size measurements are robust.

In addition to the relative internal and external error estimates, we use  991 Hectospec stellar spectra in the F2 field to estimate a lower limit on the size measurements. Almost all of stars in this sample (886 or $\sim90\%$) are detected in the HSC image and have associated ``size" measurements. The median radius measured along the major axis of the best-fit S\'ersic profiles for 90\% these stars (after exclusion of $\gtrsim1\farcs5$ outliers that are all saturated stars) is $0\farcs12$. We use this angular size as the lower limit on radius measurements along the S\'ersic profile major axis for the sample galaxies. 

Figure~\ref{f3} shows the  redshift distribution of measured major axis radii in physical units (colored circles). As expected, redshift ranges containing prominent galaxy overdensities (e.g., at $z\sim0.3$) show the broadest range of measured radii. Transformation of the angular size limit of $0\farcs12$ into physical units is a function of redshift (black solid line in Figure~\ref{f4}), and $\sim0.7\%$ of F2-HSC galaxies have size measurements below this limit. In the quiescent sample we analyze in Sections~\ref{globquiescent} and \ref{ec} $\sim1\%$ (39) of galaxies have size measurements below the limit set by stellar surface brightness profiles. Images of these objects show compact galaxy profiles. The exclusion of this very small fraction of objects does not change our results; we continue to use all size measurements in our analysis.

\begin{deluxetable*}{ccccc}
\tabletypesize{\small}
\tablecaption{Structural Prooperties of  F2-HSC Galaxies\label{table1}}
%\tablewidth{2.5in}
\tablehead{
\colhead{SHELS ID} & \colhead{R$_{mag}$} & \colhead{$z_{spec}$} & \colhead{$r_e\, [\arcsec]$} & \colhead{$b/a$}
}
\colnumbers
\startdata 
138.7221239+30.9168767 & 18.508$\pm$0.001 & 0.39824$\pm$0.00009 & 1.675$\pm$0.010 & 0.725$\pm$0.003 \\
138.7334206+30.9311798 & 19.972$\pm$0.003 & 0.33800$\pm$0.00025 & 1.040$\pm$0.009 & 0.744$\pm$0.007 \\
138.7119248+30.9047620 & 18.945$\pm$0.002 & 0.27294$\pm$0.00013 & 1.44$\pm$0.01 & 0.855$\pm$0.005 \\
138.7169235+30.9454838 & 20.262$\pm$0.004 & 0.34122$\pm$0.00010 & 0.867$\pm$0.007 & 0.956$\pm$0.009 \\
138.7369170+30.9735458 & 19.407$\pm$0.002 & 0.30834$\pm$0.00014 & 1.207$\pm$0.002& 0.301$\pm$0.005 \\
\nodata & \nodata &\nodata &\nodata &\nodata\\
\enddata
\tablecomments{This table is available in its entirety in a machine-readable form in the online journal. A portion is shown here for guidance regarding its form and content.}

\end{deluxetable*}

\subsection{Completeness}\label{com}

\begin{figure*}
\begin{centering}
%\hspace*{-0.35in}
\includegraphics[scale=0.325]{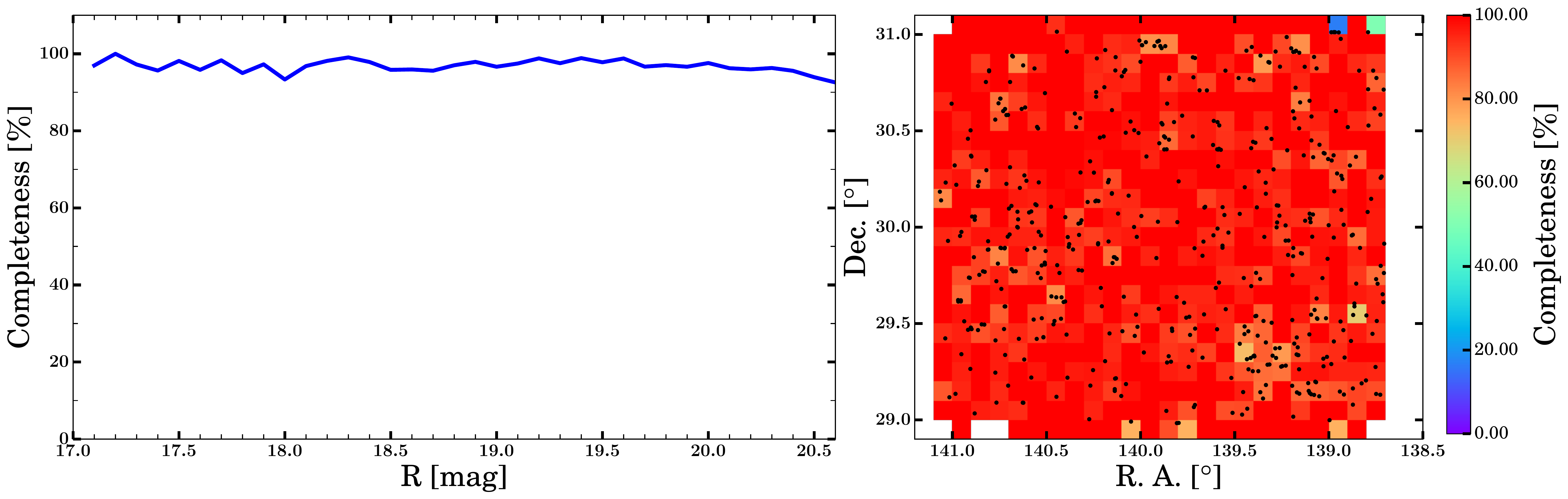}

\caption{Completeness of the SHELS F2 HSC survey in size measurements. Differential completeness as a function of $R-$band magnitude for $R_\mathrm{mag}<20.6$ F2 galaxies (left). Spectroscopic completeness in $6\arcmin\times6\arcmin$ bins for SHELS F2 HSC galaxies with $R_\mathrm{mag}<20.6$ (right).  Black points show F2 galaxies without an HSC size measurement. \label{f4}}
\end{centering}
\end{figure*}

\begin{figure}
\begin{centering}
%\hspace*{-0.35in}
\includegraphics[scale=0.35]{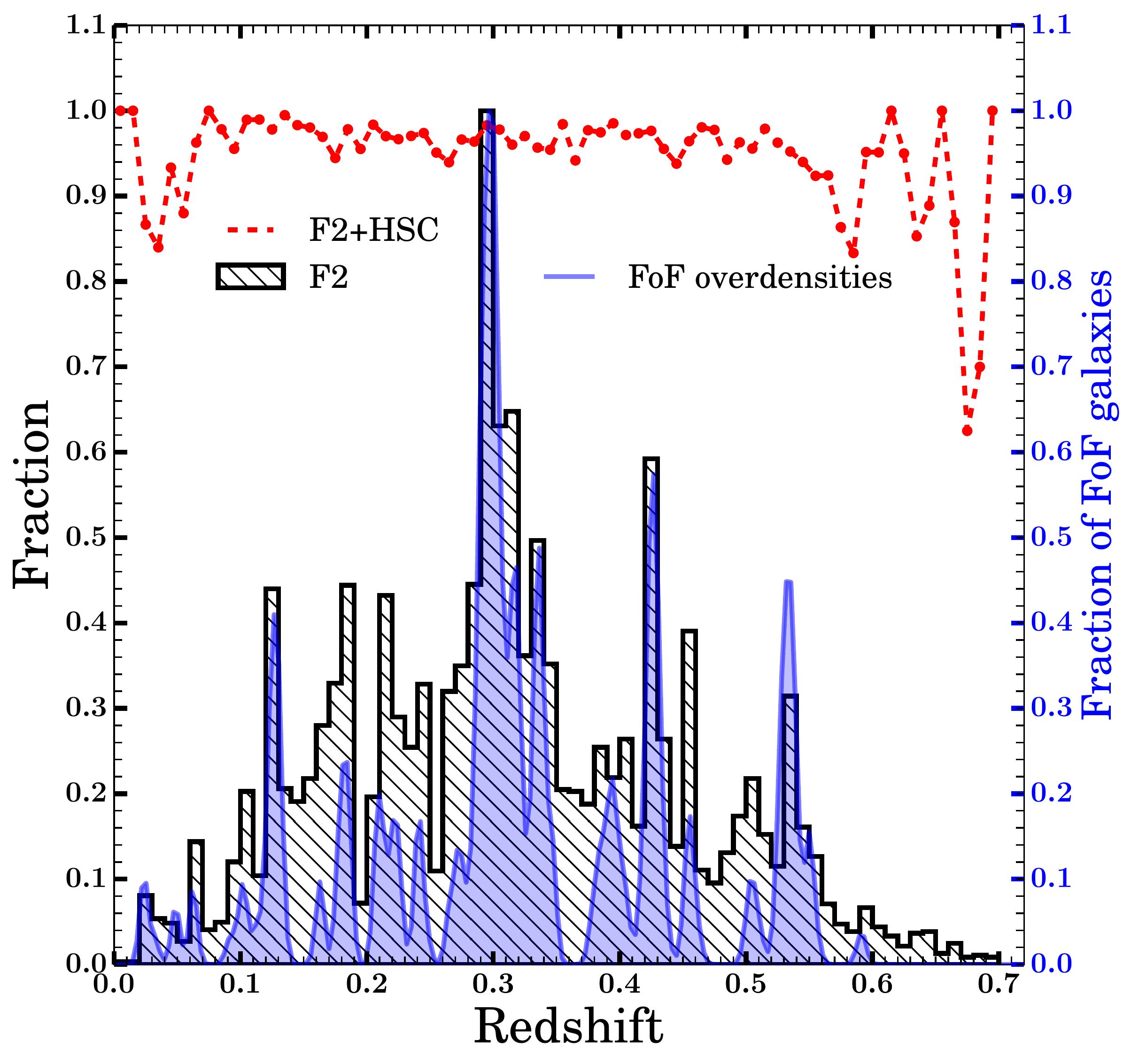}
\caption{Redshift distribution of SHELS F2 galaxies (black histogram) and the fraction of galaxies with size measurements in each redshift bin (red points and line). Blue peaks show the distribution of overdensities in the F2 field based on an FoF algorithm \citep[][see text for details]{Sohn2018}.\label{f5}}
\end{centering}
\end{figure}

To examine the completeness of the spectro-photometric sample, we first calculate the fraction of SHELS F2 galaxies with size measurements in bins of their $R-$band magnitude (left panel of Figure~\ref{f4}). For the  $R_\mathrm{mag}<20.6$ sample the differential completeness is consistently high ($\gtrsim92\%$). The SHELS F2 sample is $>90\%$ complete in almost every ($89\%$) $6\arcmin\times6\arcmin$ spatial bin of the F2 field footprint (the right panel of Figure~\ref{f4}). 

Figure~\ref{f5} shows (in red) the redshift distribution of the size measurement completeness (defined as the fraction of SHELS F2 galaxies with HSC-based sizes). Figure~\ref{f3} shows that the range of size measurements is not biased by the local galaxy number density. Comparison with the redshift distribution of SHELS F2 galaxies (black histogram of Figure~\ref{f5}, normalized to one in the most populated redshift bin) confirms that the fraction of galaxies with measured sizes is independent of structure in the F2 field. A small number of redshift bins where the size measurement completeness falls below $90\%$ are concentrated near the redshift limit of the survey where the number of spectroscopic targets  declines sharply and and thus a small number of galaxies with missing sizes represents a large fraction of galaxies in the redshift bin.

To select redshift bins that minimize the impact of peculiar velocities associated with dense structures in F2, we apply a friends-of-friends (FoF) algorithm to identify the densest regions  \citep[][and the references therein]{Sohn2018}. We adopt a standard FoF approach that connects neighboring galaxies with projected spatial  and radial  separation less than the specified values \citep{Huchra1982, Geller1983}. The projected separation between two galaxies is  $\Delta D_{ij} =  {\rm tan} (\theta_{ij}) (D_{ij}$) where $\theta_{ij}$ is the angular separation of the pair and $D_{ij}$ is the average comoving distance to the pair. The radial comoving separation is ($\Delta V_{ij}= \left|D_{c, i} - D_{c, j}\right|$) where $D_{c,i}$ and $D_{c,j}$ are the comoving distances to the individual galaxies. 

At the median redshift of the survey, $z = 0.31$, the fiducial average projected separation of galaxies in the F2 spectroscopic sample is 4.7~Mpc. The average projected separation is a function of redshift primarily as a result of the magnitude limit. At each redshift we take the limiting pairwise projected separation, $\Delta D_{ij}$, equal to 0.1 of the average projected separation of galaxies at that redshift; we take $\Delta V_{ij} = 5\times\Delta D_{ij}$. 
This procedure yields centers of FoF systems. Here, we consider only rich systems with more than 10 FoF members systems in a cylinder of radius $R_{cl}=2$~Mpc and extent $\left|\Delta(cz)/(1+z_{cl})\right|<1500$~km~s$^{-1}$ centered  on the FoF position in redshift space.  We use the resulting redshift distribution of F2 clusters and their members (blue smooth histogram in Figure~\ref{f5}) only to select the positions and widths of intervals for the investigation of evolutionary trends in  Section~\ref{nde}. The distribution of clusters is normalized to one at) the peak of distribution ($z\sim0.3$). We smooth the distribution of overdensities using a Gaussian kernel with standard deviation $\sigma_z\sim0.003$.

Some HSC extended objects lack a spectrum. For these objects, we first estimate the $i-$ band magnitude corresponding to the limiting $R-$band magnitude of the spectroscopic sample ($R_{lim}=20.6$). Following the procedure outlined in Section 2.1. of \citet{Geller2016}, we fit the $i-$band magnitudes of F2 galaxies in the HSC images (SExtractor parameter \texttt{MAG\_AUTO}) with a linear combination of their $R-$band magnitude and Sloan $r-i$ colors. The zero-point of the resulting relation gives $i_{lim}\approx R_{lim}-0.18$. To this limiting $i-$band magnitude there are $\sim6800$ HSC objects without a a spectrum and within the total footprint of the F2 field. 

Visual inspection shows that most of these objects are saturated stars. Less than 25\% ($\sim1500$) of HSC objects without spectra appear extended. Half of these extended objects are  in \citet{Geller2014} either as galaxies without a spectrum or as galaxies with a spectrum but fainter than the F2 spectroscopic completeness limit ($R>20.6$). The remaining $\sim750$ extended objects without spectroscopic counterparts  are located either at the edges of the F2 field, where the spectroscopic survey becomes sparse, or in the masked regions listed in Table~1 of \citet{Geller2014}. These regions of the F2 field are masked because the DLS photometry in those regions is unreliable. The fraction of the F2-HSC photometric sample with $i_{lim}\sim20.42$ ($\approx R_{lim}=20.6$) and without a spectra in the complete  (unmasked) region of the F2 field is negligible.

\subsection{The Size Distribution}\label{sizedist}
\begin{figure*}
\begin{centering}
%\hspace*{-0.35in}
\includegraphics[scale=0.4]{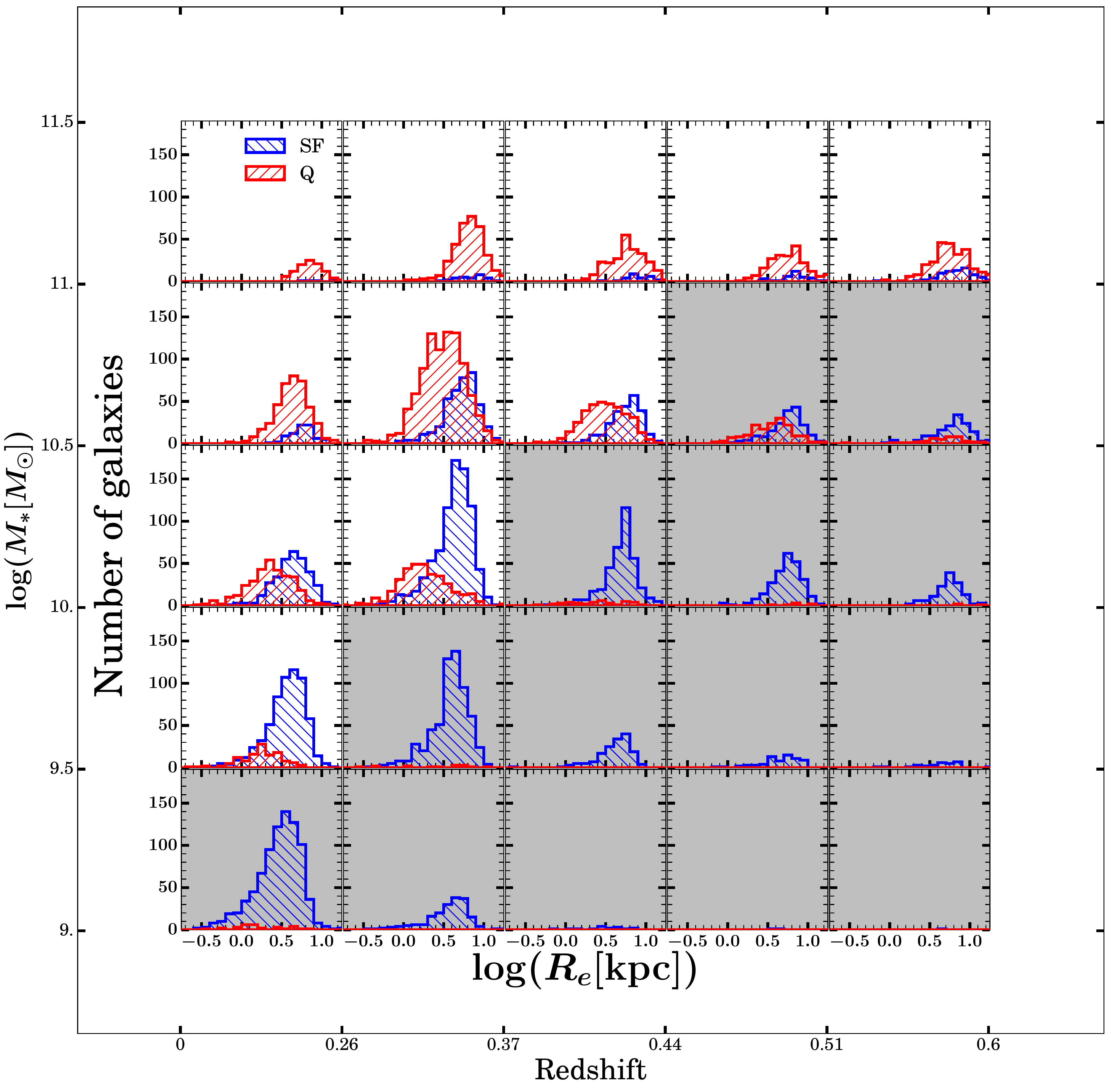}

\caption{Distribution of major axis radii for quiescent (red histograms) and star-forming (blue histograms) SHELS F2 galaxies in bins of redshift and stellar mass. We select redshift bins to encompass similar volumes (except for the first and the last redshift bin) and to include the large structures in the field. Panels with a white background represent redshift intervals where the F2-HSC sample is complete for $>85\%$ of the corresponding mass range. \label{f6}}
\end{centering}
\end{figure*}

Measurements based on imaging in broad-band filters ranging from the visible to near-infrared show that the size distributions of star forming and quiescent galaxies differ over a broad redshift range \citep[e.g., ][]{Franx2008,Williams2010,Wuyts2011,vanderWel2014,Allen2015,Straatman2015,Yano2016,Brennan2016,Faisst2017,Haines2017,Gu2018}. For $\log(M_*/M_\sun)>10$ and at $0<z\lesssim3$ star forming galaxies appear larger than quiescent systems matched in stellar mass.

The F2-HSC sample probes this difference as a function of  stellar mass and redshift. We select star forming and quiescent galaxies based on the spectroscopic indicator D$_n4000$. (Section~\ref{spec}) and map the distribution of their sizes (measured along major axis) onto a stellar mass - redshift grid (Figure~\ref{f6}). 

The magnitude limit restricts the number of populated bins and the completeness of both star forming and quiescent subsamples in each bin (see Section~\ref{masslim} for more details on the limiting mass for the quiescent subsample). The distributions of sizes for star forming (blue histograms in Figure~\ref{f6}) and quiescent systems (red histograms in Figure~\ref{f6}) in the redshift bins that are
$> 85\%$ mass complete  (panels with white background in Figure~\ref{f6})\footnote{In Section~\ref{masslim} we derive redshift-dependent mass limit for the mass-complete quiescent F2-HSC sample. Dominated by old stellar populations, quiescent galaxies have higher mass-to-light ratios than the star forming systems \citep[e.g,][] {vonderLinden2010,Ilbert2013,Geller2014}. Thus the red dashed line in Figure~\ref{f7} represents the limiting stellar mass as a function of redshift for a mass-complete sample of quiescent {\it and} star forming galaxies in the field.} show the same qualitative trend: star forming galaxies are on average larger than quiescent systems of the same mass. This trend remains if we replace the radius along major axis $R_e$ with the circularized radius $R_{e,c}$.

Samples of massive galaxies ($\log(M_\ast/M_\sun)>10)$ at $0<z<3$, divided into broad redshift intervals of $\Delta z=0.5$, display a stark difference in size distributions between the star-forming and quiescent populations with similar stellar masses \citep[e.g.,][]{vanderWel2014}. For $0.1<z<0.6$ the F2-HSC survey divided into $\Delta \log(M_*/M_\sun)=0.5$  stellar mass bins {\it and} $\Delta z\sim0.1$  redshift bins, star forming systems span a narrower range of sizes and have, on average, more extended surface brightness profiles than their massive quiescent counterparts.

\section{Properties of the Quiescent Population}\label{globquiescent}

We focus on 4210 F2-HSC  quiescent galaxies (i.e with D$_n4000>1.5$).  Based on the  limiting stellar mass as a function of redshift (Section~\ref{masslim}) we select redshift-dependent mass-complete quiescent samples and investigate the relationships among stellar mass, size, and the  stellar population age as indicated by D$_n4000$ (Section~\ref{size-mass}). We use these data along with the complete redshift survey to trace evolutionary trends for quiescent galaxies as a function of stellar mass, size, and stellar population age (Section~\ref{ec}). 

\subsection{Stellar Mass Limit}\label{masslim}

\begin{figure}
\begin{centering}
%\hspace*{-0.35in}
\includegraphics[scale=0.4]{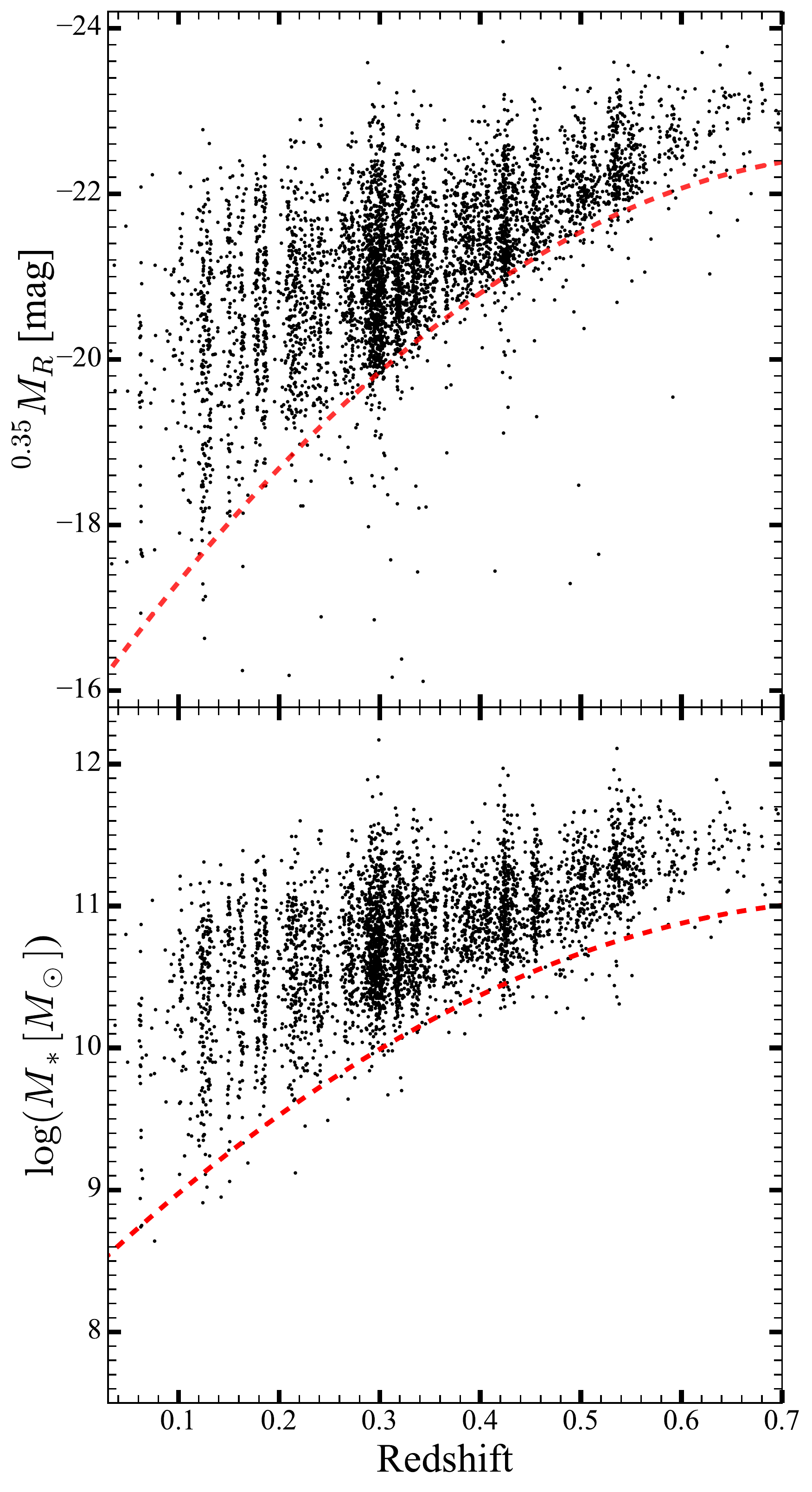}
\caption{$K-$corrected absolute R-band magnitude versus redshift (top) and stellar mass versus redshift (bottom). The points show the F2 quiescent galaxies. The red curves show the redshift dependence of the limiting absolute magnitude (top) and galaxy stellar mass  (bottom) corresponding to the magnitude limit of the survey ($R_{lim} = 20.6$).\label{f7}}
\end{centering}
\end{figure}

\begin{figure*}
\begin{centering}
%\hspace*{-0.35in}
\includegraphics[scale=0.325]{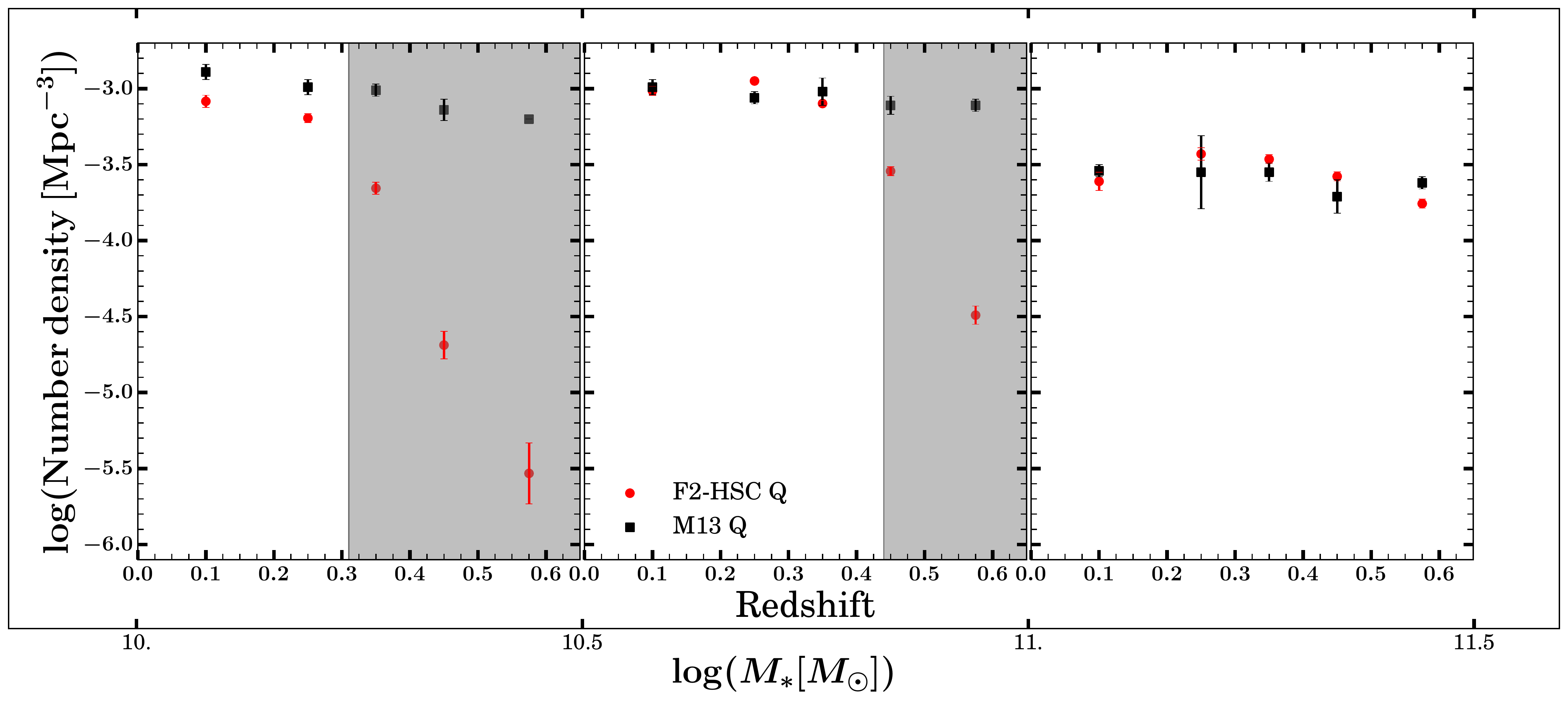}

\caption{Number density of quiescent galaxies (red symbols) as a function of redshift in bins of stellar mass compared with the number density evolution of galaxies in the PRIMUS survey \citep[][black simbols]{Moustakas2013}. In the white area the quiescent F2-HSC sample is  $>99\%$ complete at the lower stellar mass limit of the bin (see the lower panel of Figure~\ref{f7}). \label{f8}}
\end{centering}
\end{figure*}

In magnitude limited samples like SHELS F2 the limiting stellar mass where the galaxy sample is complete reflects the observable distribution of absolute magnitudes as a function of redshift. The upper panel of Figure~\ref{f7} shows the absolute $R-$band magnitude as a function of redshift for F2-HSC galaxies with D$_n4000>1.5$. We determine the $K-$corrected magnitudes using the SDSS $ugriz$ photometry and the {\texttt kcorrect} code \citep{Blanton2007}. The red dashed line (upper panel) traces the limiting absolute magnitude ($K-$corrected to the approximate mean redshift $z\sim0.35$ of the quiescent F2 sample), $^{0.35}M_{R,lim}$, as a function of measured redshift:

\begin{equation}
^{0.35}M_{R,lim}=m_{R,lim}-5\log\left(\frac{D_L(z)}{10\, \mathrm{pc}}\right)-\overline{K}(z),
\end{equation}

\noindent where $m_{R,lim}=20.6$~mag is the limiting apparent magnitude for the complete sample, $D_L(z)$ is the luminosity distance, and $\overline{K}(z)$ is the average $K$~correction (to $z=0.35$) for quiescent (D$_n4000>1.5$) galaxies as a function of galaxy redshift. A combination of photometric errors and a large scatter around the average $K$~correction results in a small fraction of quiescent galaxies (288, $\sim6\%$) that fall below the calculated absolute magnitude limit. 

In the lower panel of Figure~\ref{f7} we show the distribution of stellar masses  as a function of redshift. We transform the magnitude limit $^{0.35}M_{R,lim}$ into the galaxy stellar mass limit $M_{\ast,lim}$ (red dashed line in the lower panel) using the mass-to-light ratio $(M/L)_R\sim2$ based on the approximately constant $(M/L)_R$ value for quiescent (D$_n4000>1.5$) systems in the SHELS F2 galaxy sample \citep[see Figure~12 of][]{Geller2014}. Eighty-eight quiescent galaxies ($<2\%$ of the complete quiescent sample) lie below the calculated stellar mass limit as a result of photometric errors.

Figure~\ref{f8} compares the evolution in number density of the F2-HSC quiescent sample  with the abundances of quiescent galaxies in the  5.5~deg$^2$ (distributed over five separate fields) grism survey, PRIMUS \citep{Coil2011,Moustakas2013}. The quiescent PRIMUS sample is selected based on galaxy position in the star-formation rate vs. stellar mass diagram. Although the quiescent sample selection criteria are not identical, differences between the number densities in the F2-HSC and PRIMUS samples are small; the densities are within $\pm 2\sigma$ for essentially all mass bins and at all redshifts. The only significant difference occurs for galaxies with $10<\log(M_\ast/M_\sun)<10.5$ in the redshift interval $0.2<z<0.3$. The upper limit of this redshift interval, $z\sim0.3$, is at the completeness limit for F2 galaxies with lower stellar masses (down to $\log(M_\ast/M_\sun)=10$, Figure~\ref{f7}). Furthermore, the $0.2<z<0.3$  redshift range includes a large void at $z\sim0.25$   \citep[Figure~\ref{f5}, see also Figure~6 of][]{Geller2014}.  Based on the median redshift transverse length, aspect ratio, and the median radial depth of the survey \citep{Driver2010}, the expected cosmic variance for the F2 field is 10\% \citep{Geller2016}. The overall similarity between the F2 and PRIMUS surveys in Figure~\ref{f8} demonstrates that F2 is large enough to measure quiescent galaxy number densities that are insensitive to cosmic variance at the level relevant for this study.

\subsection{Size -- Stellar Mass -- D$_n4000$ Correlation}\label{size-mass}

\begin{figure*}
\begin{centering}
%\hspace*{-0.3in}
\includegraphics[scale=0.5]{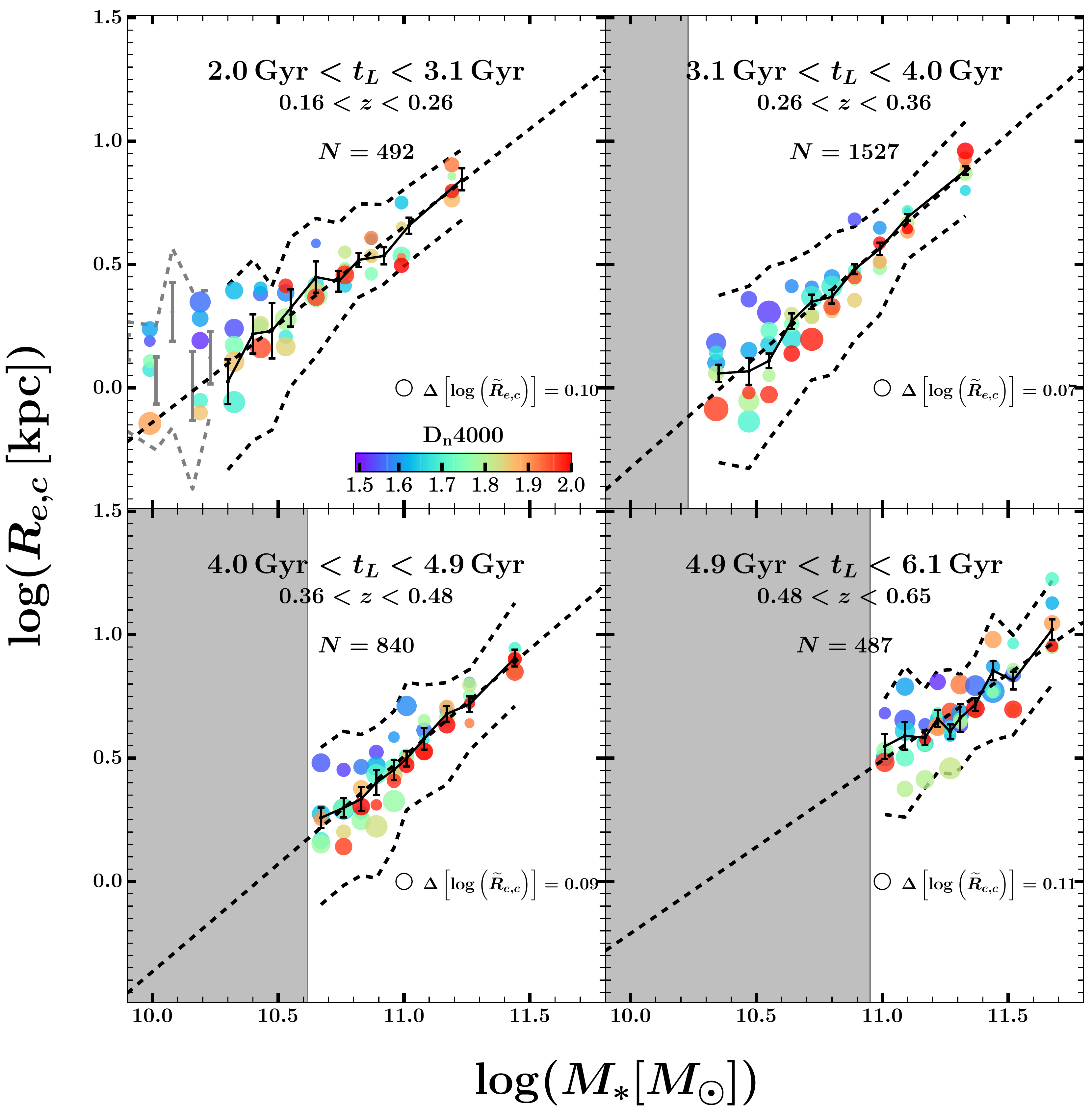}
\caption{Median circularized half-light radius as a function of stellar mass and D$_n4000$  for the quiescent F2-HSC galaxies in four similar lookback time intervals. The dot color indicates the median D$_n4000$ 
in the bin.  The dot size is normalized to the typical bootstrapped error for the median size in each redshift interval. The legend gives the value of the typical error in the median size and the number of  galaxies ($N$). The white regions show stellar mass ranges where the quiescent  sample is  $>99\%$ complete. In each panel the solid black line connects the median sizes in 10 equally populated mass bins. The errors in median values are bootstrapped. The region between the broken dashed lines shows the intrinsic scatter and includes 68\% of the sample in each mass bin. The central black dashed line traces the best-fit linear relation between stellar mass and size. The gray error bars and gray broken dashed lines in the first panel indicate the median size (with bootstrapped errors) and the intrinsic scatter in the stellar mass range where the relation between stellar mass and size flattens. We exclude these points from the fit.\label{f9}}
\end{centering}
\end{figure*}

The size distribution of quiescent galaxies  depends strongly on galaxy stellar mass (red histograms in Figure~\ref{f6}). Figure~\ref{f9} shows quiescent galaxies in  stellar mass - galaxy size  space. The four intervals  correspond to similar lookback time intervals. For each subsample of quiescent galaxies, we  trace the median galaxy sizes in equally populated stellar mass bins (error bars in Figure~\ref{f9}) and in  intervals that include the central 68\% of the population in each mass bin (broken dashed lines in Figure~\ref{f9}). We analyze only the complete mass bins (i.e., white regions in the plot). 

Table~\ref{table2} lists the parameters defining  the  relation $R_e[\mathrm{kpc}]=A\times\left(M_\ast/10^{11}\, M_\sun\right)^\alpha$ that best describes median trend (black dashed lines in Figure~\ref{f9}; we fit the relation using~\texttt{Python} function \texttt{scipy.optimize.curve\_fit}). Our  normalization of the size -- stellar mass relation (median size of an $M_\ast=10^{11}\, M_\sun$ galaxy) ensures that the corresponding mass bins are populated across the redshift range that we probe ($0.1<z<0.6$). To fit the size -- stellar mass relation for $0.16<z<0.26$ we exclude galaxies with stellar mass $M_\ast\lesssim2\times 10^{10}\, M_\sun$, \citep[][ grey error bars and grey dashed lines in the first panel of Figure~\ref{f9}]{vanderWel2014} where the relation flattens as a result of observational selection.

The size -- stellar mass relation in Figure~\ref{f9} confirms known  trends: a) the slope of the linear relation in log-log space ($\alpha$) remains constant (within $\pm 2\sigma$ ) over the full redshift interval $0.16<z<0.65$ and b) the zero point of the relation ($A$) shifts to smaller median sizes for $M_\ast=10^{11}\, M_\sun$ quiescent systems as the redshift increases. The form of the size -- stellar mass relation depends on  the range of stellar masses, errors in the size and stellar mass measurements, the resolution of the images, surface brightness selection effects, the imaging rest-frame wavelength, and the range of galaxy environments  \citep[e.g,][]{Lange2015,Sweet2017}.  Nevertheless, the F2-HSC relations are fully consistent with the slope parameters derived for similarly selected quiescent galaxy samples  \citep[$0<z<1$,][]{vanderWel2014}\footnote{\citet{vanderWel2014} derive  the stellar mass -- size relation using radii along the galaxy major axis. With the same  size definition, the relation for F2-HSC quiescent galaxies has a slope $\alpha$ that ranges from $0.60\pm0.05$ for $0.16<z<0.26$ to $0.84\pm0.03$ for $0.36<z<0.48$, and is (within $\pm 2\sigma$) consistent with the results for the 3D-HST/CANDLES sample at $z\sim0.25$}, in the local universe \citep[SDSS sample,][]{Guo2009}, and at higher redshifts \citep[$1<z<1.5$,][]{Newman2012}. 

The extensive spectroscopy provides the age indicator D$_n$4000 which is an additional  dimension for exploring the formation history of the quiescent population.  As in \citet{Zahid2017}, for each redshift range in Figure~\ref{f9} we segregate objects into ten equally populated stellar mass bins and divide these bins further into five equally populated bins in D$_n4000$. For every  bin we compute the median galaxy size (circularized half-light radius) and bootstrap the errors. 

Quiescent galaxies in the local volume covered by the SDSS have decreasing size with increasing average stellar population age at each stellar mass (\citealt{Zahid2017}, see also \citealt{Wu2018}). The F2-HSC quiescent sample is an order of magnitude smaller than the local SDSS sample. However, Figure~\ref{f9} demonstrates that the anti-correlation between quiescent galaxy size and D$_n4000$ index is present at intermediate redshift. The median size (normalized by stellar mass)  decreases with increasing D$_n4000$  over the redshift range $0.16<z<0.48$ for stellar masses $M_\ast\lesssim10^{11}\, M_\sun$.  Sparse sampling at $z>0.48$ over a narrow stellar mass range  precludes extension of the relation to greater redshift.  As in the local universe, the  size--stellar mass relations are essentially parallel  as a function of D$_n4000$  (straight dashed black lines in Figures~\ref{f9}) in the most populated redshift intervals (second and third panel of Figure~\ref{f9}). This self-similarity is  important  for  characterizing  quiescent galaxies selected at fixed size or at sizes normalized by the trend with stellar mass (Section~\ref{nde}).

Quiescent F2-HSC galaxies share global properties with other samples of similarly massive quiescent galaxies at $0\lesssim z\lesssim1$ (Section~\ref{globquiescent}). The  number densities are approximately redshift independent for $M_\ast>10^{10}\, M_\sun$ quiescent galaxies. The slope of the mass-size relation is also constant ($\alpha\sim0.85$).  However, the zero point (i.e., the average size of a quiescent galaxy with $M_\ast=10^{11}\, M_\sun$) increases by $\sim45\%$ between $z\sim0.55$ and $z\sim0.2$. The anti-correlation between quiescent galaxy size and its D$_n4000$ (a proxy for average stellar population age) at constant stellar mass extends from $z\sim0$ to $z\sim0.5$. The results complement analyses of quiescent galaxy evolution based on the SDSS and on samples at greater redshift. 

\begin{deluxetable*}{cccc}
\tabletypesize{\small}
\tablecaption{Best-fit Parameters of the Quiescent Galaxy Size -- Stellar Mass Relation:  $R_{e,c}[\mathrm{kpc}]=A\times\left(\frac{M_\ast}{10^{11}\, M_\sun}\right)^\alpha$\label{table2}}
%\tablewidth{2.5in}
\tablehead{
\colhead{Redshift range} & \colhead{Stellar mass range\tablenotemark{a}}& \colhead{$\log(A)$} & \colhead{$\alpha$}
}
\colnumbers
\startdata 
$0.16<z<0.26$& $10.24<\log(M_\ast/M_\sun)<11.6$\tablenotemark{b} & $0.65\pm 0.02$ & $0.79\pm0.05$\\
$0.26<z<0.36$ & $10.24< \log(M_\ast/M_\sun)<12.17$ & $0.58\pm 0.02$ & $0.90\pm0.04$\\  
$0.36<z<0.48$ & $10.62 <\log(M_\ast/M_\sun)< 11.97$ & $0.506 \pm 0.007$ & $0.87 \pm 0.03$\\
$0.48<z<0.65$ & $10.96 <\log(M_\ast/M_\sun)< 12.11$  & $0.49 \pm 0.03$ & $0.70 \pm 0.09$\\
\enddata
\tablecomments{
\tablenotetext{a}{The lower-mass limit is the mass completeness limit at the upper limit of the redshift range (lower panel in Figure~\ref{f7}).}
\tablenotetext{b}{The lower-mass limit eliminates the stellar mass range where the size -- stellar mass relation flattens \citep[e.g.,][]{vanderWel2014}.}
}
\end{deluxetable*}

\section{Evolutionary Constraints}\label{ec}

\subsection{The Size--Mass Relation}\label{smr}

\begin{figure*}
\begin{centering}
%\hspace*{-0.35in}
\includegraphics[scale=0.4]{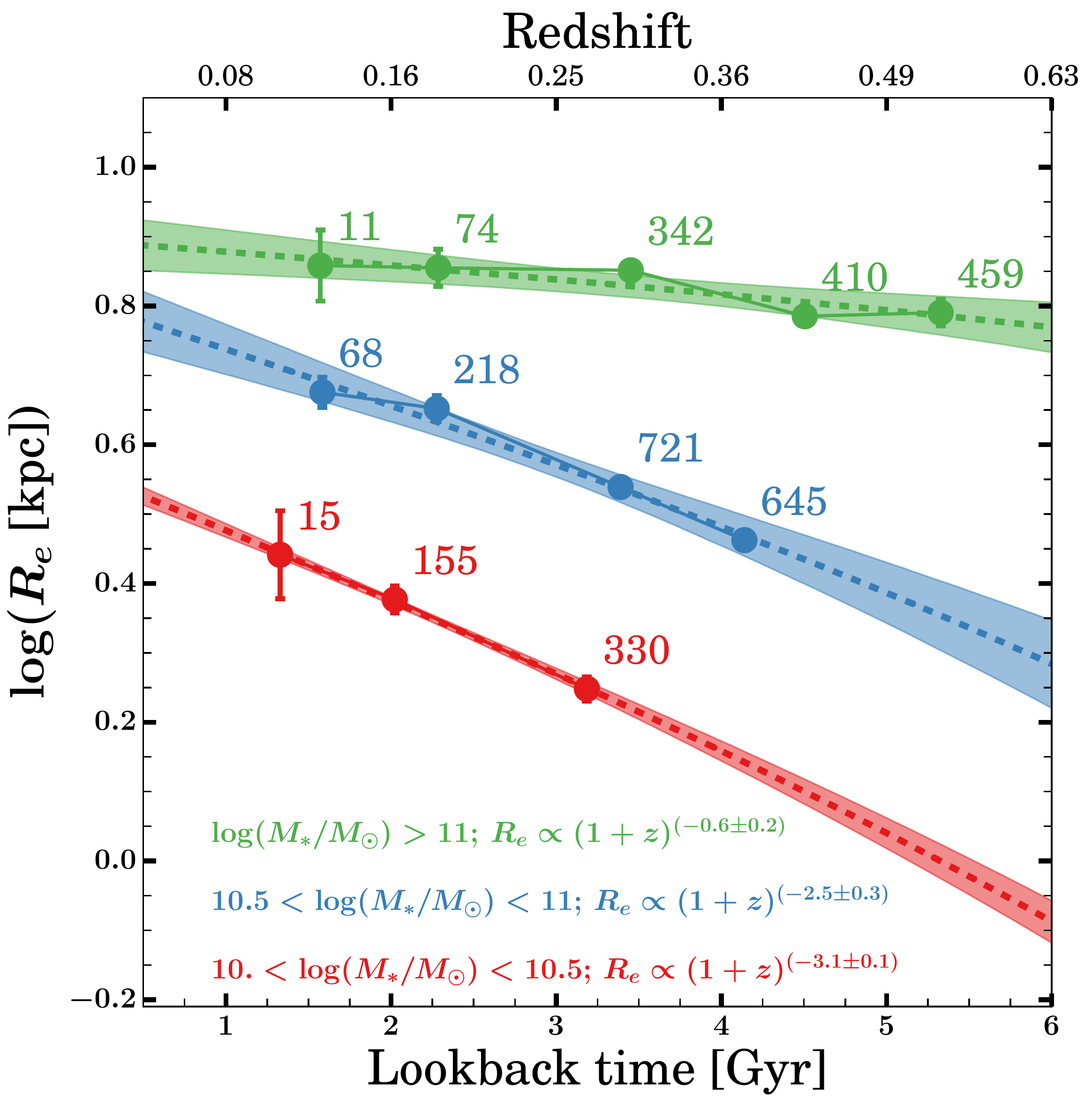}
\caption{Redshift evolution of the median size for quiescent F2-HSC galaxies in three stellar mass bins. Filled circles and associated numbers correspond to the median size, median lookback time (redshift), and the number of  mass-selected quiescent systems in time intervals of $\lesssim1$~Gyr (described in more detail in the caption of Figure~\ref{f17}). The number of time/redshift intervals where we trace the evolution in the median galaxy size is limited by the highest redshift where galaxy sample in a given stellar mass bin is complete (Figure~\ref{f7}). Dashed lines and shaded regions are the best-fit relation $R_e=B\times(1+z)^\beta$ and its 95\% confidence interval, respectively. Table~\ref{table3} lists the median sizes and values of the parameters $\log(B)$~and~$\beta$ for the three stellar mass intervals.\label{f10}}
\end{centering}
\end{figure*}

\begin{figure*}
\begin{centering}
%\hspace*{-0.35in}
\includegraphics[scale=0.4]{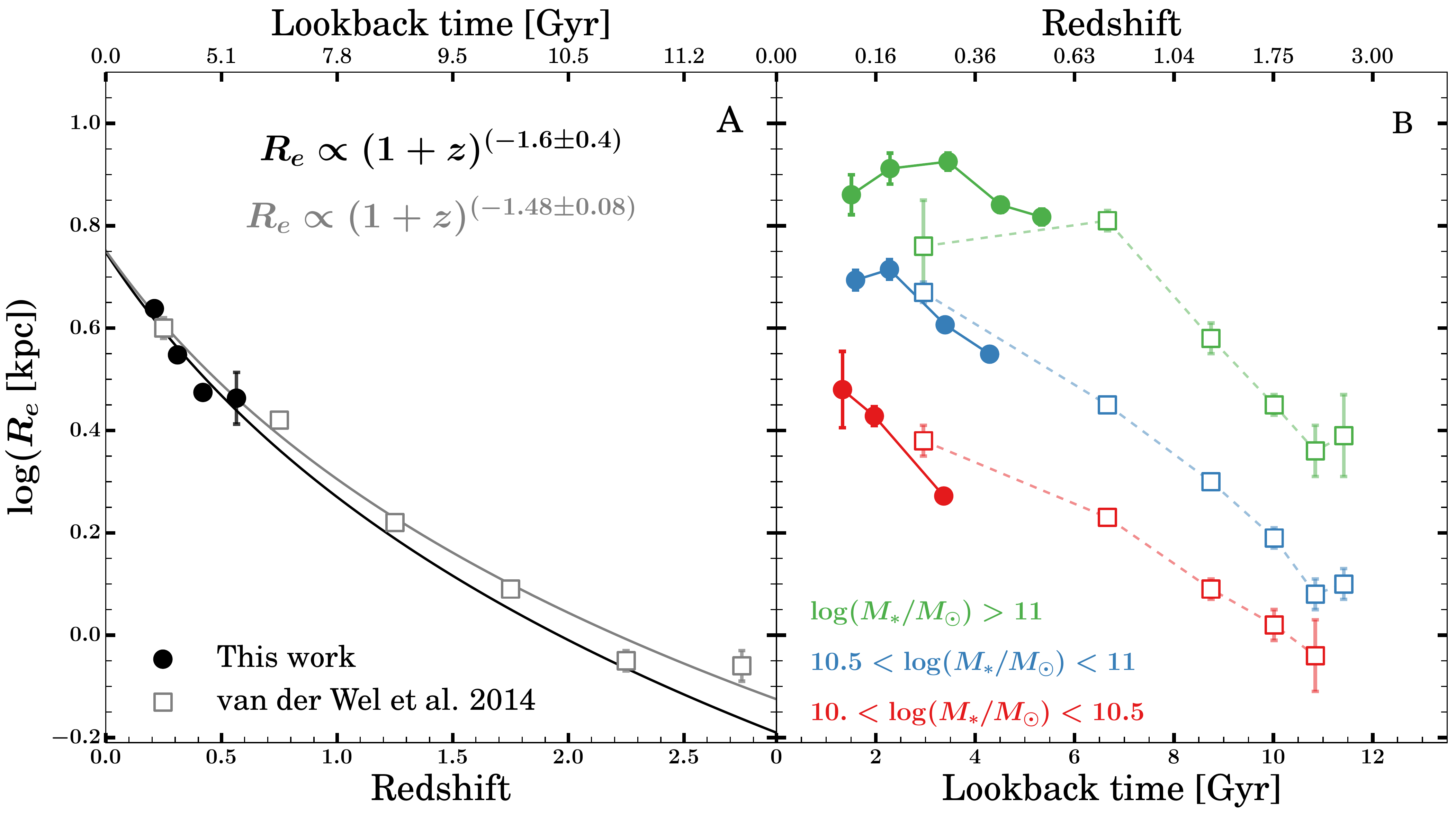}
\caption{A: Redshift evolution of the median size for an $M_\ast=5\times10^{10}\, M_\sun$ quiescent galaxy (black solid line) compared to the median size evolution of 3D-HST galaxies of the same mass \citep[gray solid line,][]{vanderWel2014}. B: Redshift evolution of the median size for quiescent galaxies in three stellar mass bins; filled circles  indicate  F2-HSC galaxies and open squares represent the 3D-HST sample. We account for systematics in stellar mass estimates by shifting the stellar masses of F2-HSC galaxies to match the two curves in panel A at $z=0$. Note the consistency of the two samples in panel A. \label{f11}}
\end{centering}
\end{figure*}

\begin{deluxetable*}{ccccccc}
\tabletypesize{\small}
\tablecaption{Median size evolution of quiescent F2-HSC galaxies\tablenotemark{a}\label{table3}}
%\tablewidth{70pt}
\tablehead{
\multicolumn{7}{c}{\normalsize  $R_e=B\times(1+z)^\beta$}\\
\hline
 \colhead{}& \multicolumn{6}{c}{\normalsize Stellar mass range\tablenotemark{b}}\\
\colhead{} & \multicolumn{2}{c}{$[10-10.5]\, M_\sun$} & \multicolumn{2}{c}{$[10.5-11]\, M_\sun$} & \multicolumn{2}{c}{$[>11]\, M_\sun$}\\
\colhead{\normalsize Redshift} &\colhead{\normalsize $\log(B)$} &\colhead{\normalsize $\beta$} &\colhead{\normalsize $\log(B)$}  &\colhead{\normalsize $\beta$}&\colhead{\normalsize $\log(B)$} &\colhead{\normalsize $\beta$}\\
&&&&&&\\
\hline
\colhead{0.09--0.31} &\colhead{$0.575\pm0.008$} & \colhead{$-3.1\pm0.1$} &\colhead{} & \colhead{}  &\colhead{} & \colhead{} \\
\colhead{0.09--0.44} &\colhead{} & \colhead{} & \colhead{$0.68\pm0.02$} &\colhead{$-2.5\pm0.3$} &\colhead{} & \colhead{} \\
\colhead{0.09--0.60} &\colhead{} & \colhead{}  &\colhead{} & \colhead{} &\colhead{$0.90\pm0.02$} &\colhead{$-0.6\pm0.2$}\\
&&&&&&\\
\hline
\hline
\multicolumn{7}{c}{\normalsize $log(\widetilde{R_e}\, \mathrm{[kpc]})$}\\
\hline
\colhead{}& \multicolumn{6}{c}{\normalsize Stellar mass range}\\
\colhead{\normalsize Redshift\tablenotemark{b}}& \multicolumn{2}{c}{\normalsize $[10-10.5]\, M_\sun$} & \multicolumn{2}{c}{\normalsize $[10.5-11]\, M_\sun$} & \multicolumn{2}{c}{\normalsize $[>11]\, M_\sun$}\\
}
\startdata 
0.09--0.11 & \multicolumn{2}{c}{$0.44\pm0.07$} &  \multicolumn{2}{c}{\nodata} &  \multicolumn{2}{c}{\nodata}\\
0.11--0.20 & \multicolumn{2}{c}{$0.38\pm0.02$} &  \multicolumn{2}{c}{\nodata} &  \multicolumn{2}{c}{\nodata}\\
0.20--0.31 & \multicolumn{2}{c}{$0.34\pm0.04$} &  \multicolumn{2}{c}{\nodata} &  \multicolumn{2}{c}{\nodata}\\
0.09--0.13 & \multicolumn{2}{c}{\nodata} &  \multicolumn{2}{c}{$0.68\pm0.02$} &  \multicolumn{2}{c}{\nodata}\\
0.13--0.22 & \multicolumn{2}{c}{\nodata} &  \multicolumn{2}{c}{$0.65\pm0.02$} &  \multicolumn{2}{c}{\nodata}\\
0.22--0.32 & \multicolumn{2}{c}{\nodata} &  \multicolumn{2}{c}{$0.54\pm0.01$} &  \multicolumn{2}{c}{\nodata}\\
0.32--0.44 & \multicolumn{2}{c}{\nodata} &  \multicolumn{2}{c}{$0.46\pm0.02$} &  \multicolumn{2}{c}{\nodata}\\
0.09--0.14 & \multicolumn{2}{c}{\nodata} &  \multicolumn{2}{c}{\nodata} &  \multicolumn{2}{c}{$0.86\pm0.06$}\\
0.14-0.23 & \multicolumn{2}{c}{\nodata} &  \multicolumn{2}{c}{\nodata} &  \multicolumn{2}{c}{$0.85\pm0.03$}\\
0.23--0.34 & \multicolumn{2}{c}{\nodata} &  \multicolumn{2}{c}{\nodata} &  \multicolumn{2}{c}{$0.85\pm0.02$}\\
0.34--0.46 & \multicolumn{2}{c}{\nodata} &  \multicolumn{2}{c}{\nodata} &  \multicolumn{2}{c}{$0.79\pm0.01$}\\
0.46--0.60& \multicolumn{2}{c}{\nodata} &  \multicolumn{2}{c}{\nodata} &  \multicolumn{2}{c}{$0.79\pm0.02$}\\
\enddata
\tablecomments{
\tablenotetext{a}{Median sizes in this table correspond to the F2-HSC data in Figure~\ref{f10} (filled circles). The parameters $\log(B)$ and $\beta$ describe best-fit relations between the half-light radius $R_e$ and redshift for quiescent F2-HSC systems segregated by stellar mass.}
\tablenotetext{b}{Redshift intervals correspond to $\lesssim1$~Gyr time intervals;  the highest redshift interval starts at the completeness limit for each stellar mass range.  These redshift/time intervals are equivalent to the ones we describe in Section~\ref{gm} and use in Figure~\ref{f17}.}
}
\end{deluxetable*}

For massive ($M_\ast>10^{10}\, M_\sun$) quiescent galaxies, the slope of the mass-size  relation is constant for $ 0.1< z < 0.6$, but the normalization decreases significantly as the redshift increases (Section~\ref{size-mass}, Figure~\ref{f9}). This change in the size--mass relation zero point traces the evolution in the typical (circularized) size of an $M_\ast=10^{11}\, M_\sun$ quiescent galaxy (Table~\ref{table2}). Using the size  along the major axis ($R_e$) instead of $R_{e,c}$ does not alter the constancy of the slope of the size -- stellar mass relation. We mostly follow trends in $R_e$  to make direct comparisons with results at higher redshifts. The results are independent of the definition of galaxy size.

To explore the dependence of the  size growth of quiescent galaxies on the stellar mass range, we divide the F2-HSC quiescent sample into three stellar mass bins spanning the $10^{10}\, M_\sun\leq M_\ast\lesssim 8\times10^{11}\, M_\sun$ range. For each subsample we calculate the median size (with bootstrapped errors) in redshift bins that correspond to lookback time intervals of $\lesssim1$~Gyr starting from the lookback time/redshift completeness limit for each mass-segregated subsample (as described in Section~\ref{gm}). Table~\ref{table3} lists the redshift intervals and corresponding median sizes of the mass-selected quiescent galaxies.

The redshift evolution of a typical F2-HSC galaxy size depends on its stellar mass (filled circles in Figure~\ref{f10}). The parameters of the $R_e=B\times(1+z)^\beta$ relation (listed in the top section of Table~\ref{table3}) confirm that $10<\log(M_\ast/M_\sun)<10.5$ galaxies grow most rapidly in size: their growth rate over the redshift interval $0.1<z\lesssim0.3$ is $|\beta|\equiv|d\log(R_e)/d\log(1+z)|=3.1\pm0.1$, higher than the $|\beta|$ values for more massive quiescent systems ($|\beta|=2.5\pm0.3$ for $10.5<\log(M_\ast/M_\sun)<11$ and $|\beta|=0.6\pm0.2$ for $M_\ast>10^{11}\, M_\sun$).  The growth rate of  the least massive galaxies in our sample ($10<\log(M_\ast/M_\sun)<10.5$) is consistent (within  $\pm2\, \sigma$ ) with the $|\beta|$ for $10.5<\log(M_\ast/M_\sun)<11$ quiescent galaxies. The most massive ($M_\ast>10^{11}\, M_\sun$ ) F2-HSC systems do grow significantly more slowly ($\lesssim25\%$ over the $0.1<z<0.6$ redshift interval) than the least massive quiescent galaxies that increase their size by $\gtrsim70\%$ between $z\sim0.3$ and $z\sim0.1$. The difference between growth rates for these two mass-selected quiescent subsamples is at $\sim10\, \sigma$ level.

We  match the zero point of the mass-size relation to the median size of a $M_\ast=5\times10^{10}\, M_\sun$ galaxy and compare the exponential size growth of this fiducial galaxy in the F2-HSC and 3D-HST \citep{vanderWel2014} surveys in Figure~\ref{f11}A. Because the two samples derive stellar masses using different methods, we apply an offset of $\Delta[\log(M_\ast/M_\sun)]=-0.11$~dex to our stellar mass estimates to match the $R_e(z=0)$ points of the two relations. We then examine the differences in the trends in size with redshift. Although the F2-HSC sample spans a very different  redshift range than 3D-HST, the exponents of the two $R_e-z$ relations agree within $\pm(1-1.5)\, \sigma$. The F2-HSC $0.1<z<0.6$ sample thus confirms that the typical size of a $5\times10^{10}\, M_\sun$ quiescent galaxy evolves with redshift as $\log(R_e)\propto(1.5-1.6)\times \log(1+z)$.

The median sizes of mass-selected quiescent galaxies at $z\geq0.25$ based on HST imaging \citep[open squares in Figure~\ref{f11}B,][]{vanderWel2014} also agree well with the HSC-based results (filled circles in Figure~\ref{f11}B) in the overlapping redshift interval ($z\sim0.25$) for galaxies in the stellar mass range $10^{10}\, M_\sun<M_\ast<10^{11}\, M_\sun$. The most massive ($11<\log(M_\ast/M_\sun)<11.5$) 3D-HST galaxies at $z<0.5$ have slightly smaller sizes than their massive F2-HSC counterparts. The origin of this discrepancy is probably  the smaller volume probed by 3D-HST at $z<0.5$(Section~\ref{cv}). The largest galaxies are rare and the volume of 3D-HST may be insufficient to contain them.

To test the impact of survey volume on the number and average size of the most massive quiescent galaxies we scale the number of massive ($M_\ast>10^{11}\, M_\sun$) galaxies in F2 to the number expected in 
0.25~degree$^{2}$ survey area \citep[i.e., the total area of the 3D-HST survey, ][]{Momcheva2016}. With the reduced area, cosmic variance increases to $25\%$ \citep{Driver2010} and we include this factor in the absolute uncertainty of the scaled massive galaxy counts ($39\pm12$). We draw 10,000 samples of F2 galaxies with a total number of galaxies that (approximately) follows a Gaussian distribution with $\mu=39$ and $\sigma=12$. The median value and $[16,85]\%$ interval of the distribution of median sizes in the simulated massive galaxy samples - $\log(R_e[kpc])=0.88^{+0.05}_{-0.04}$ - is within $1\sigma$ from the median size of $M_\ast>10^{11}\, M_\sun$ 3D-HST galaxies at $z\sim0.25$ \citep[$0.76\pm0.09$, ][]{vanderWel2014}. Thus the small volume of the  3D-HST survey at $z<0.5$ completely accounts for the difference between the median sizes of the most massive intermediate-redshift galaxies in the two surveys (green circles and green square at $z=0.25$ in Figure~\ref{f11}B).

The mass-dependence of the trends in size with redshift at $z>0.5$ differs from the trends we observe at lower redshift: more massive galaxies grow relatively more rapidly  at $0.5<z<3$  \citep[open squares in Figure~\ref{f11}B,][]{vanderWel2014}. The F2-HSC survey  shows that at intermediate redshift  the more rapid size growth shifts from the most massive  ($M_\ast>10^{11}\, M_\sun$) to the least massive systems with $M_\ast\sim10^{10}\, M_\sun$. Over the $\sim6$~Gyr of lookback time ($0.5<z<3$) probed by the 3D-HST survey, the most massive galaxies grow quickly; $10^{10}\, M_\sun<M_\ast<3\times10^{10}\, M_\sun$ galaxies expand only moderately. In contrast, over the $\sim5$~Gyr of lookback time ($z<0.5$) probed by the F2-HSC quiescent systems with $M_\ast>10^{11}\, M_\sun$ grow slowly; the size growth of $10^{10}\, M_\sun<M_\ast<3\times10^{10}\, M_\sun$ galaxies accelerates. In other words, size growth for 0.1 $ < z < 0.6$ is not a simple extrapolation of the growth observed at $z \gtrsim 0.5$.

\subsection{D$_n$4000}\label{dn}
\begin{figure*}
\begin{centering}
%\hspace*{-0.35in}
\includegraphics[scale=0.425]{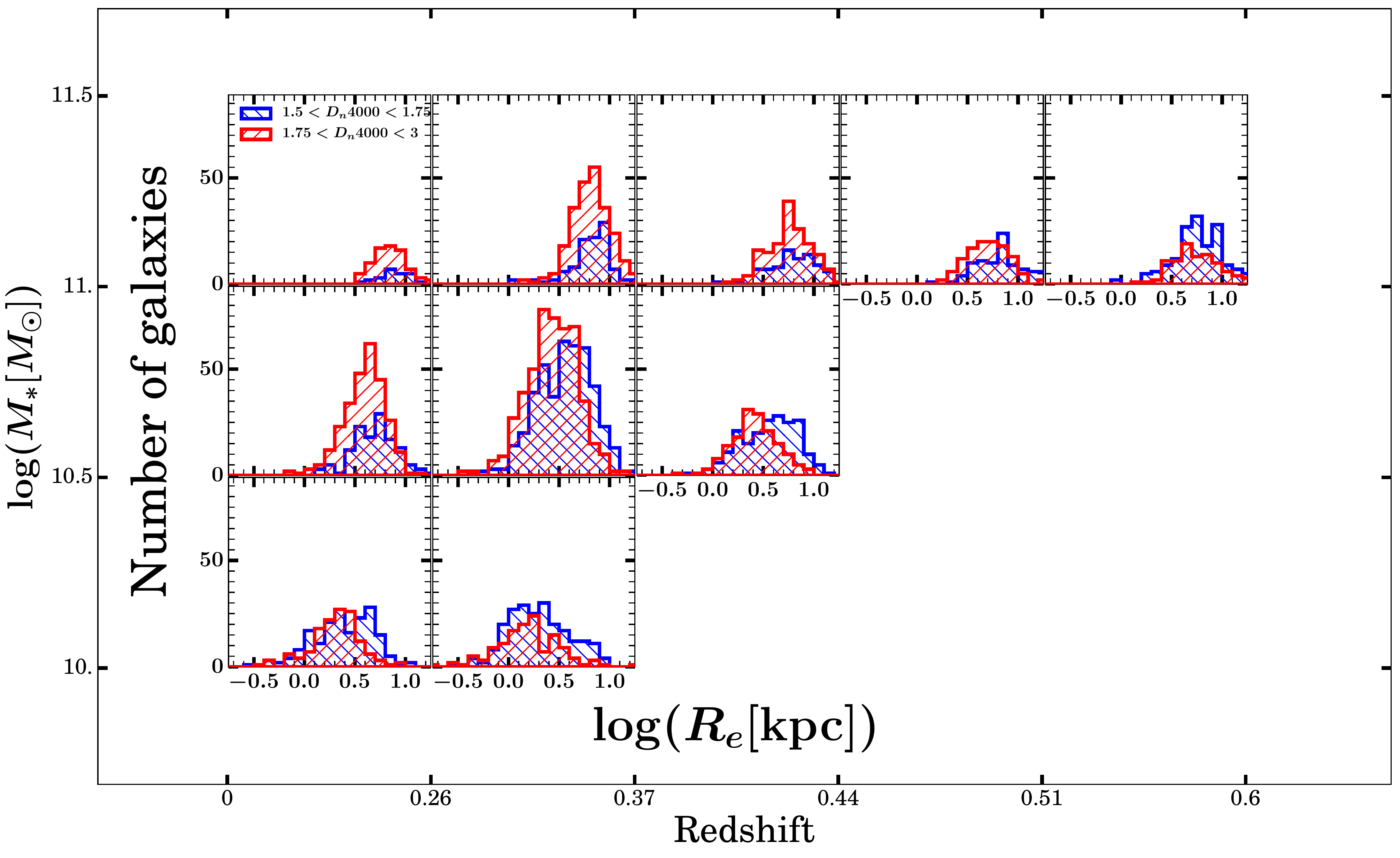}

\caption{Distribution of major axis radii for quiescent F2-HSC galaxies segregated by redshift and stellar mass and further divided into subsamples of low (blue histograms) and high (red histograms) D$_n4000$. We divide the samples at the median D$_n$4000 for the parent sample. The redshift bins  encompass similar volumes (except for the first and the last redshift bin) and  include large structures in the field. In each stellar mass bin (row) we show distributions only in the redshift intervals where the F2-HSC sample is complete for $>85\%$ of the mass range. \label{f12}}
\end{centering}
\end{figure*}

\begin{figure*}
\begin{centering}
%\hspace*{-0.35in}
\includegraphics[scale=0.4]{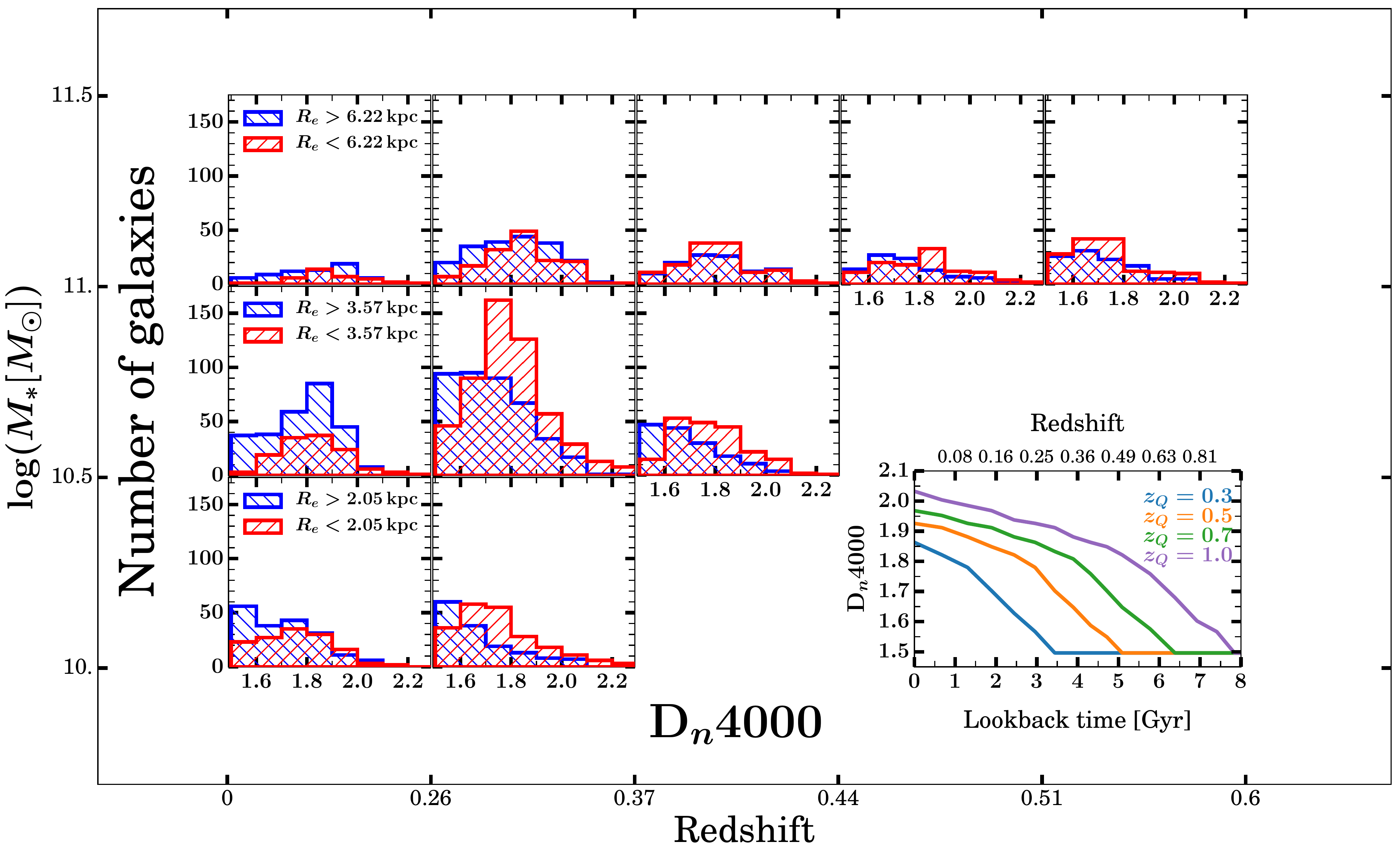}

\caption{Large Matrix: Distribution of D$_n4000$ indices for quiescent F2-HSC galaxies segregated by redshift and stellar mass and further divided into two subsamples at the median size for galaxies in each  stellar mass bin. Blue and red histograms indicate larger and smaller galaxy sizes, respectively. The selection of redshift bin intervals and complete subsamples are the same as in Figure~\ref{f12}. Inset: D$_4000$ evolution of model galaxies that cease forming stars at a range of redshifts: $0.3\leqslant z_Q\leqslant1$. See text for details. \label{f13}}

\end{centering}
\end{figure*}

Relative to star forming systems, the size distribution of quiescent  F2-HSC galaxies is clearly offset towards smaller average sizes for stellar masses $M_\ast<10^{11}\, M_\sun$ at all redshifts  where both the star forming and quiescent samples are complete (Section~\ref{sizedist}, Figure~\ref{f6}). To further probe the relation between galaxy size and age indicator  D$_n4000$ , we investigate 1) size and 2) D$_n4000$ distributions for subsamples segregated by stellar mass and redshift. We explore the sensitivity of the size distributions to D$_n$4000 in stellar mass bins as a function of redshift. We then explore the sensitivity of the D$_n$4000 distributions in stellar mass and redshift bins to galaxy size. 

Quiescent $M_\ast<10^{11}\, M_\sun$ galaxies at $z<0.5$ dominated by younger stellar populations ($1.5<$D$_n4000<1.75$, blue histograms in the second and third row of Figure~\ref{f12}) are on average larger than similarly massive quiescent systems in the same redshift intervals but with older stellar populations (D$_n4000>1.75$, red histograms in the second and third row of Figure~\ref{f12}). With very low $p-$values ($\lesssim10^{-3}$), both Kolmogorov-Smirnov (K-S) and Anderson-Darling (A-D) 2-sample tests confirm that in all stellar mass--redshift cells the $M_\ast<10^{11}\, M_\sun$ size distributions of ``young" and ``old" quiescent galaxies are not drawn from the same parent distribution. Conversely, in only one redshift cell containing the most massive ($M_\ast>10^{11}\, M_\sun$) quiescent F2-HSC galaxies ($0.44<z<0.51$) K-S and A-D tests provide $p-$values as low as $\sim2\times10^{-3}$. The size distributions of the most massive quiescent systems ($M_\ast >10^{11} M_\sun$) in all other redshift intervals with $z<0.6$ are consistent with being drawn from the same underlying distribution in D$_n4000$. 

Distributions of D$_n4000$ for quiescent systems in stellar mass -- redshift bins separated into pairs of subsamples based on galaxy size relative to the median size for their stellar mass range (Figure~\ref{f13}) confirms the results in Figure~\ref{f12}. At $z<0.5$ and $M_\ast<10^{11}\, M_\sun$ galaxies with smaller sizes (red histograms in Figure~\ref{f13}) have on average higher D$_n4000$ (i.e., older stellar population) than more extended galaxies sharing the same redshift and stellar mass  bin (blue histograms in Figure~\ref{f13}). Statistical K-S and A-D tests confirm that pairs of D$_n4000$ distributions in stellar mass -- redshift cells with $M_\ast<10^{11}\, M_\sun$ do not originate from the same parent population. For the massive ($M_\ast>10^{11}\, M_\sun$) quiescent subsamples K-S and A-D once again provide similarly low $p-$values as for lower-mass systems only in the $0.44<z<0.51$ range, where the number of galaxies is lower than in all other subsamples with low $p-$values. Other massive quiescent systems show no significant difference in the distribution of stellar population ages for the two size bins.

The inset of Figure~\ref{f13} shows the time evolution of the D$_n4000$ index for a galaxy model that estimates the period of quiescence (i.e, the average age) for F2-HSC systems in the stellar mass -- redshift matrix of Figure~\ref{f13}.  We measure D$_n4000$  from the synthetic spectrum of a quiescent galaxy constructed using the Flexible Stellar Population Synthesis code \citep[FSPS;][]{Conroy2009, Conroy2010}. The model galaxy has solar metallicity and and constant star formation rate for 1~Gyr ending at redshift $z_Q$. The average age of a quiescent galaxy is then the difference between the lookback time at the redshift of a galaxy  and the lookback time corresponding to the redshift of quiescence $z_Q$ for the model curve that matches measured the D$_n4000$ value at the galaxy redshift.  

The age of  $M_\ast>10^{11}\, M_\sun$ galaxies does not change with galaxy size. The majority of galaxies populating the first row in Figure~\ref{f13} span the age range from $\sim2$~Gyr at $z\sim0.55$ to $\sim4.5$~Gyr at $z\lesssim0.26$. For $M_\ast<10^{11}\, M_\sun$ systems, the estimated average age depends both on galaxy stellar mass {\it and} size. For the majority of larger intermediate-mass systems (blue histograms in the second row of Figure~\ref{f13}) the age varies from $\lesssim1$~Gyr at $z\sim0.3$ to $\sim3$~Gyr at $z\sim0.15$.  However, at $z>0.37$ intermediate mass systems ($10.5<\log(M_\ast/M_\sun)<11$) with larger sizes have an average age  between these two limits ($\sim2$~Gyr). A large fraction of smaller galaxies of the same mass (red histograms in the second row of Figure~\ref{f13}) have similar ages across this redshift range ($\sim3$~Gyr). The majority of $10<\log(M_\ast/M_\sun)<10.5$ quiescent systems with larger sizes (blue histograms in the third row of Figure~\ref{f13}) are consistently very young ($<1$~Gyr) at $z<0.37$. A significant fraction of their smaller counterparts (red histograms in the third row of Figure~\ref{f13}) have older ages ranging  from $\sim1$~Gyr at $z\sim0.3$ to $\sim2$~Gyr at $z\sim0.15$. 

The distributions of quiescent galaxy sizes   in Figure~\ref{f12} show that the trend of smaller sizes for older galaxies continues beyond the size difference between star forming and quiescent galaxies (Figure~\ref{f6}). Throughout the observed redshift range, quiescent $M_\ast<10^{11}\, M_\sun$ galaxies in the lower half of the age distribution are consistently larger than their similarly massive older quiescent counterparts. For $M_\ast>10^{11}\, M_\sun$ quiescent galaxies, there is no significant difference in size distributions between younger and older systems. The typical age of $M_\ast>10^{11}\, M_\sun$ quiescent galaxies divided into small and large size subsamples in different redshift intervals  (Figure~\ref{f13}) suggests passive evolution without significant size growth. At lower stellar masses the typical galaxy age depends on both galaxy stellar mass and  size (following the anti-correlation between size and age at fixed stellar mass in Figure~\ref{f9}).

\subsection{Number Density}\label{nde}

\begin{figure*}
\begin{centering}
%\hspace*{-0.35in}
\includegraphics[scale=0.4]{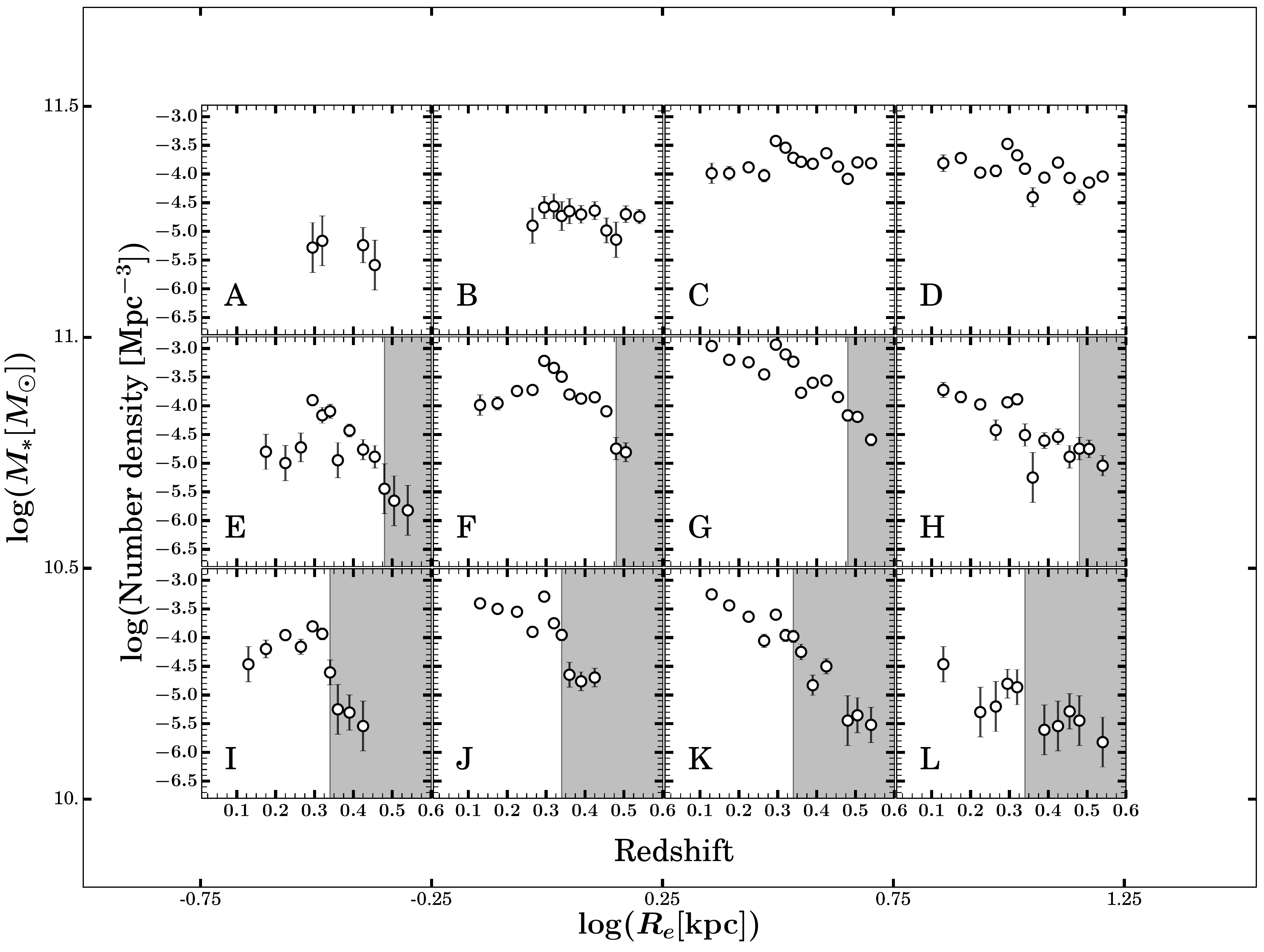}

\caption{The redshift evolution of number density for quiescent F2 galaxies in bins of mass and size. Grey shaded areas mask redshift ranges where the F2 quiescent sample is incomplete at the lower mass limit of teh bin (based on the stellar mass limit from Figure~\ref{f7}). Galaxies with sizes $-0.25<\log(R_e)<0.75$ and masses $10<\log(M_\ast)<11$ show the  most prominent evolution in number density. In contrast, the most massive (top row) galaxies display little variation in their number density with increasing redshift. For galaxies in the lowest mass bin (bottom row) the number density remains approximately constant over the redshift range where the quiescent  F2 sample is mass-complete. \label{f14}}
\end{centering}
\end{figure*}

\begin{figure*}
\begin{centering}
%\hspace*{-0.35in}
\includegraphics[scale=0.4]{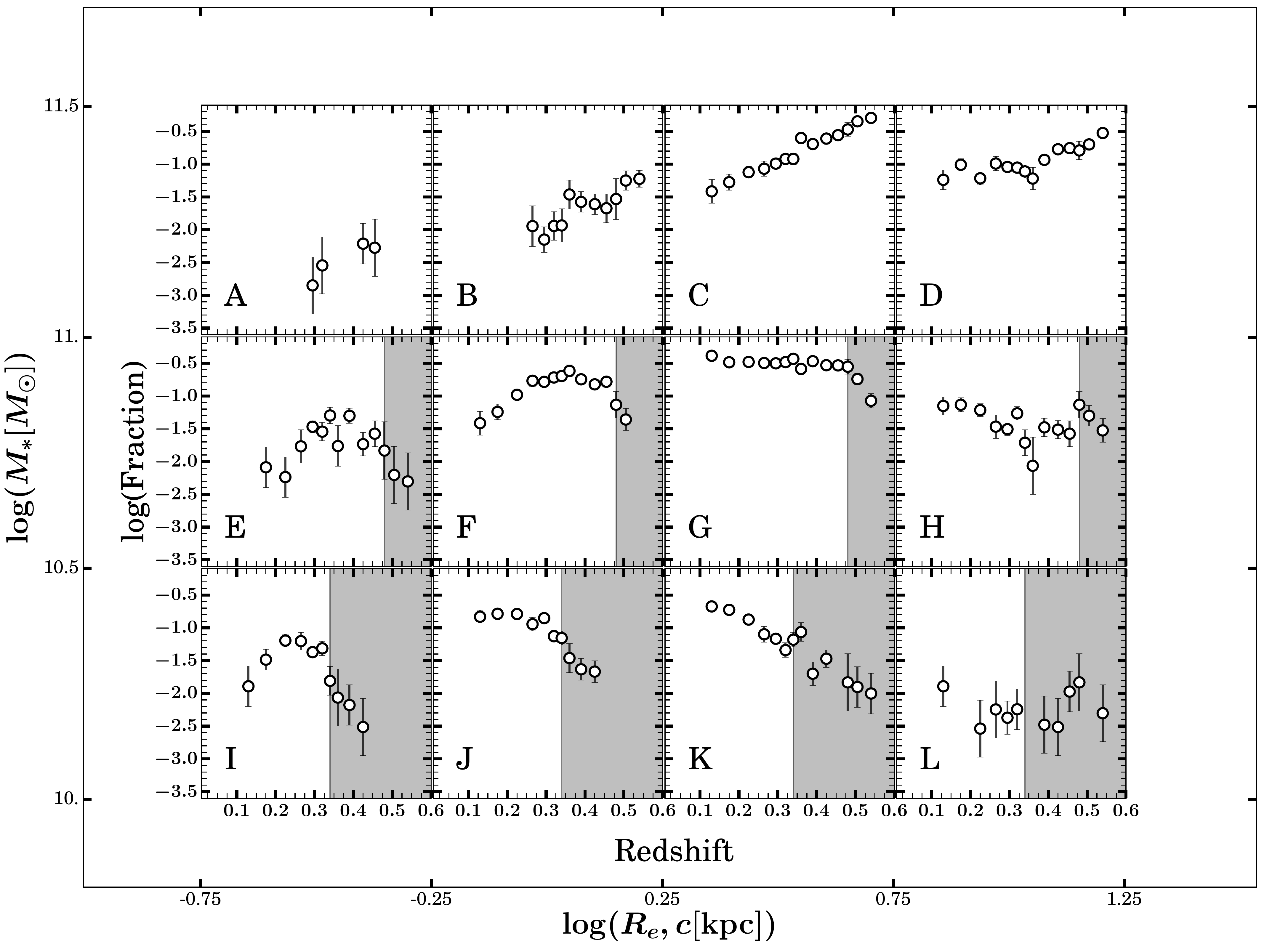}

\caption{Fractional number density evolution of quiescent F2 galaxies in bins of mass and size with respect to the underlying quiescent population. Symbols indicate the ratio between the number of quiescent F2 galaxies of a given mass and size and the total number of quiescent F2 systems in redshift bins of $\Delta z=0.05$. The first and the last redshift bin are larger: $\Delta z=0.2$ and $\Delta z=0.1$, respectively. Grey shaded areas mask redshift ranges where the F2 quiescent sample is incomplete at the lower mass limit of the bin (based on the stellar mass limit from Figure~\ref{f7}).  \label{f15}}
\end{centering}
\end{figure*} 

\begin{figure*}
\begin{centering}
%\hspace*{-0.35in}
\includegraphics[scale=0.4]{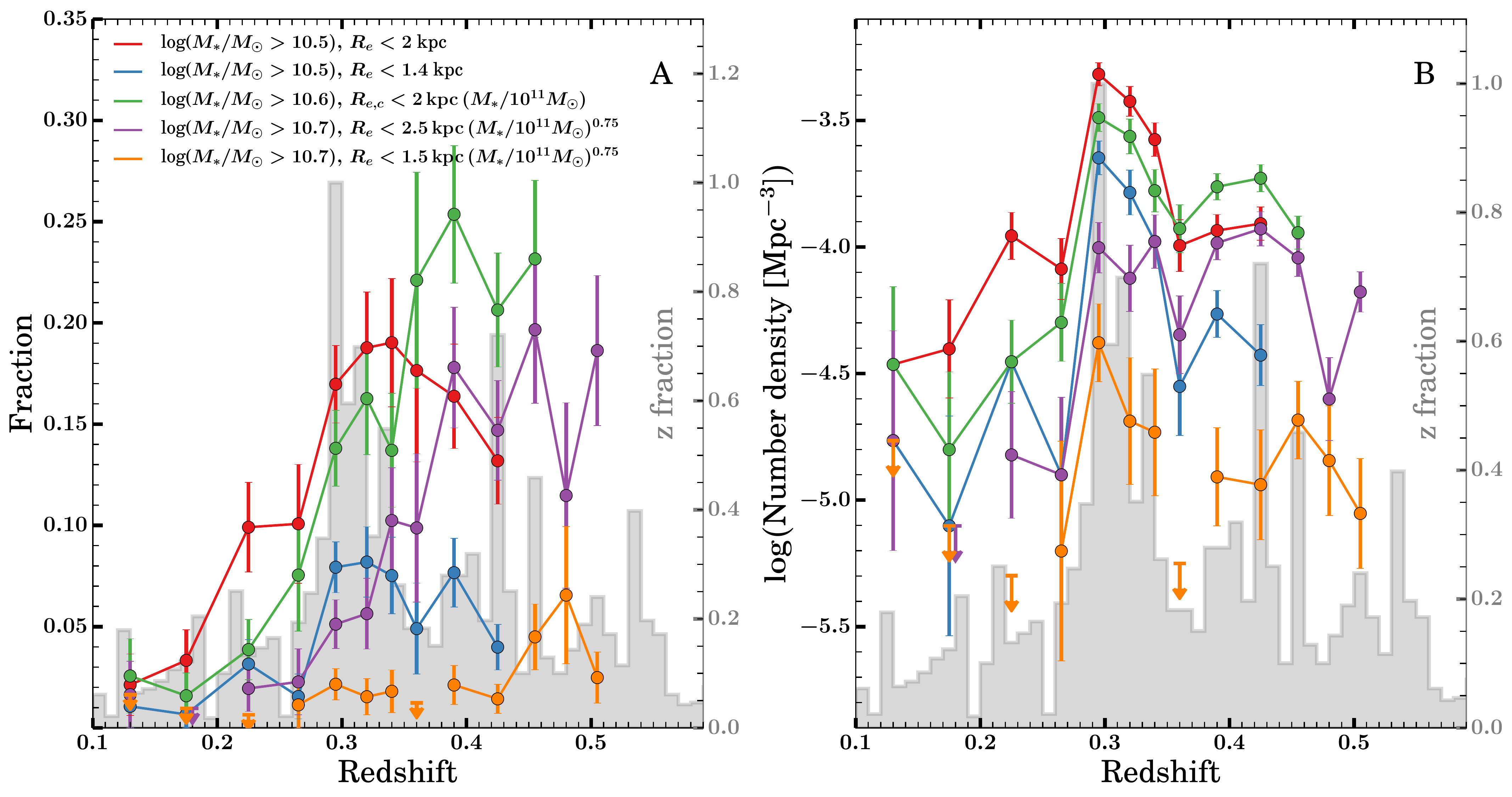}
\caption{A: Fraction of compact galaxies in equivalently massive F2-HSC quiescent subsamples as a function of redshift. B: Number density of quiescent compact F2-HSC galaxies. In both panels points are color-coded based on the definitions of compactness explored in \citet{Charbonnier2017}. Arrows indicate upper limits on fractions and number densities in redshift intervals where there are no compact systems in the field. We selected redshift intervals that follow the distribution of overdensities in the F2 field (peaks in the redshift distribution of $\log(M_\ast /M_\sun)>10.5$ quiescent systems  represented by the grey histogram). Table~\ref{table4} lists the number densities and fractions of compact massive galaxies shown here. \label{f16}}
\end{centering}
\end{figure*}

Changes in absolute and fractional abundances provides additional constraints on the processes that drive the evolution of quiescent systems. We trace the redshift evolution of galaxy number density and fraction (in the parent quiescent population) across a two-dimensional matrix in stellar mass and size (Figures~\ref{f14}~and~\ref{f15}). We measure the absolute and relative number of quiescent systems in redshift intervals defined by the distribution of  galaxy overdensities in the F2 field (blue shaded areas in Figure~\ref{f5}).

The most massive quiescent systems ($M_\ast>10^{11}\, M_\sun$) show very little change in galaxy number density with redshift (the first row in Figure~\ref{f14}). The most prominent fluctuations ($\sim0.5$~dex jumps) in the number densities of larger massive systems ($R_e\gtrsim1.8$~kpc, panels C and D in Figure~\ref{f14}) are in the redshift intervals with the largest structures in the field ($z\sim0.3$ and $z\sim0.42$). In contrast , the fraction of the most massive systems in the quiescent galaxy population declines with decreasing redshift (the first row of Figure~\ref{f15}). The trends in absolute and fractional numbers of $M_\ast>10^{11}\, M_\sun$ galaxies suggest that: 1) at the massive end of quiescent galaxy mass distribution the majority of $z\lesssim0.6$ systems do not evolve in size  individually (as the number density remains constant in every size bin), and 2) at $z\lesssim0.6$ the fraction of the most massive galaxies of all sizes within the general quiescent population declines because the properties of galaxies joining the red sequence  change with redshift.
Progressively larger numbers of less massive systems join the quiescent population with decreasing redshift \citep[downsizing,][]{Cowie1996}. 

At $M_\ast<10^{11}\, M_\sun$ the evolution in number densities and fractions of quiescent galaxies in different stellar mass and size bins becomes more complex. In the $z\lesssim0.45$ redshift interval, where the $3\times10^{10}\, M_\sun<M_\ast<10^{11}\, M_\sun$ quiescent subsample is complete, larger intermediate-mass F2-HSC galaxies ($R_e\gtrsim1.8$~kpc) have an increasing number density and are an approximately constant fraction of the parent quiescent population with decreasing redshift (panels G and H in Figures~\ref{f14}~and~\ref{f15}). As in the $M_\ast>10^{11}\, M_\sun$ quiescent subsample, substantial  departures from the monotonically increasing number densities coincide with the largest F2 structures. Smaller intermediate-mass quiescent systems depart from constant number density only in the redshift range dominated by the largest clusters in the field ($z\sim0.3$, panels E and F in Figures~\ref{f14}). In contrast, the fraction of small intermediate-mass quiescent systems decreases monotonically with decreasing redshift from $z\sim0.4$ (panels E and F in Figures~\ref{f13}). Taken together, the trends in number density and fraction show that 1) intermediate-mass quiescent galaxies experience changes in number density and fractional abundance that depend on galaxy size, 2) individual compact intermediate-mass quiescent systems, like their more massive compact counterparts, do not grow substantially in size from $z\sim0.45$ to $z\sim0.1$, and 3) the contribution of compact intermediate-mass systems to the underlying quiescent population does change with redshift suggesting additional global effects.

The evolution of number densities and fractional abundances for $10<\log(M_\ast/M_\sun)<10.5$ quiescent galaxies to $z\sim0.3$ (with stellar masses $10^{10}\, M_\sun<M_\ast<3\times10^{10}\, M_\sun$) is similar to the intermediate-mass sample. For $R_e>1.8$~kpc the number density increases towards $z\sim0.1$, and low-mass galaxies with smaller sizes show an approximately constant fractional abundance in this redshift range (modulo sharp changes related to the largest overdensity at $z\sim0.3$, last row of panels in Figure~\ref{f14}). However, unlike  quiescent galaxies with larger masses, the fraction of extended low-mass galaxies  increases with decreasing redshift (panels K and L in Figure~\ref{f15}). For smaller sizes ($R_e<1.8$~kpc), the fraction of low-mass systems is constant with redshift only for a small range in size: it declines for $R_e<0.5$~kpc (panels I and J in Figure~\ref{f15}). Evolutionary trends in absolute and fractional abundance become increasingly dependent on galaxy size with decreasing stellar mass.

Trends in the number density and fractional abundance with redshift for quiescent galaxies segregated by stellar mass and size suggest multiple channels of galaxy evolution that include internal processes
affecting the size growth of individual galaxies  and external, global processes that affect the distributions of properties for  all quiescent systems. To further examine the relative contributions of processes that drive quiescent galaxy evolution, we trace the absolute and fractional abundance of the most compact quiescent galaxies from $z\sim0.1$ to $z\sim0.6$.

The definition of a massive compact quiescent galaxy is somewhat arbitrary. The range of massive compact galaxy properties can be defined either by a constant cutoff values in stellar mass {\it and} half-light radius \citep[e.g.,][]{Trujillo2009,Carollo2013}, or by a combination of a constant lower limit on stellar mass and a lower limit on size that scales with mass \citep[a line approximately parallel to the size -- stellar mass relation at a given redshift; e.g.,][]{Valentinuzzi2010a,Taylor2010,Barro2013,Cassata2011,vanderWel2014}. We use five different criteria \citep[compiled from the literature by][]{Charbonnier2017} to select massive compact galaxies and to follow the intermediate redshift evolution of their absolute number density and of their fractional abundance in the parent quiescent population with the same stellar mass (Table~\ref{table4} and Figure~\ref{f16}).

From $z\sim0.1$ to $z \sim 0.3$, the fractional abundance of all compact samples increases  (Figure~\ref{f16}A),  \citep[Section~\ref{com}, Figure~\ref{f5},][]{Geller2014}. Interestingly, at $z\gtrsim0.3$, the location of the most massive structure in the F2 field,  the samples diverge. After the first plateau, the fraction of compact galaxies with sizes normalized by stellar mass (i.e, $\sim$ parallel to the size-mass relations of Figures~\ref{f9}) in the parent population of similarly massive quiescent systems ($\log(M_\ast/M_\sun)>10.6\, (10.7)$) continues to increase to $z\sim0.42$, and then flattens out at the redshift limit of the survey. The fractional abundance of compacts with a constant cutoff size in the quiescent sample covering a slightly broader stellar mass range ($\log(M_\ast/M_\sun)>10.5$), on the other hand, decreases for $z\gtrsim0.3$. Using a photometric sample that covers $\sim170$~deg$^2$ of the Stripe 82 field \citet{Charbonnier2017} finds the same variance between trends in fractional abundances for compact samples that include different stellar mass ranges and employ different cutoff sizes (their Figure 5).

Figure~\ref{f16}B shows similar evolutionary trends in number densities for different compact selections: compact galaxy number densities closely follow the redshift distribution of structures in the field, a trend expected if massive compact systems reside preferentially in dense environments at intermediate redshift \citep{Valentinuzzi2010a,Damjanov2015b}. Although the presence of the large overdensity at $z\sim0.3$ significantly affects the observed number density evolution, massive compact systems have at most a mild decline in number density from $z\sim0.5$ to $z\sim0.1$ of $\lesssim 0.5$~dex (or a factor of 3). 

The combined evolution in the absolute and fractional abundances of massive compact quiescent galaxies is very similar to the evolution of most the massive galaxies regardless of size (top row in Figures~\ref{f14}~and~\ref{f15}) and of the most compact intermediate-mass systems (panels E and F in Figures~\ref{f14}~and~\ref{f15}). For both the most massive and the densest galaxies, the number density tracks the distribution of structures in the field and their fraction within the general quiescent population decreases with decreasing redshift. 

\begin{deluxetable*}{ccccccccccc}
\tabletypesize{\tiny}
\tablecaption{Properties of Massive Quiescent Compact Galaxies in F2-HSC\label{table4}}
\tablewidth{70pt}
\tablehead{
 \colhead{}& \multicolumn{10}{c}{\scriptsize  Definition}\\
\colhead{} & \multicolumn{2}{c}{$R_e<2$\tablenotemark{a}} &  \multicolumn{2}{c}{$R_e<1.4$} & \multicolumn{2}{c}{$R_{e,c}<2\, (M_\ast/10^{11}M_\sun)$} 
& \multicolumn{2}{c}{$R_e<2.5\, (M_\ast/10^{11}M_\sun)^{0.75}$} & \multicolumn{2}{c}{$R_e<1.5\, (M_\ast/10^{11}M_\sun)^{0.75}$}\\ 
&&&&&&&&&&\\
\colhead{} & \multicolumn{10}{c}{\scriptsize  Stellar mass range}\\
\colhead{} & \multicolumn{2}{c}{$\log(M_\ast/M_\sun>10.5)$} & \multicolumn{2}{c}{$\log(M_\ast/M_\sun>10.5)$} & \multicolumn{2}{c}{$\log(M_\ast/M_\sun>10.6)$}  & \multicolumn{2}{c}{$\log(M_\ast/M_\sun>10.7)$} & \multicolumn{2}{c}{$\log(M_\ast/M_\sun>10.7)$}\\
&&&&&&&&&&\\
\hline
&&&&&&&&&&\\
\colhead{\scriptsize Redshift} & \multicolumn{10}{c}{\scriptsize  Number of compacts}\\
\colhead{\scriptsize 0.11--0.56} & \multicolumn{2}{c}{\scriptsize 384} & \multicolumn{2}{c}{\scriptsize 154} & \multicolumn{2}{c}{\scriptsize 480} & \multicolumn{2}{c}{\scriptsize 266} & \multicolumn{2}{c}{\scriptsize 46}\\
&&&&&&&&&&\\
\hline
\colhead{Redshift\tablenotemark{b}} & \colhead{Fraction} & \colhead{ $\log(n)$\tablenotemark{c}} & \colhead{Fraction} & \colhead{$\log(n)$} & \colhead{Fraction} & \colhead{ $\log(n)$} & 
\colhead{ Fraction} & \colhead{$\log(n)$} & \colhead{Fraction} & \colhead{ $\log(n)$}
}
\colnumbers
\startdata 
0.11--0.15 & $0.02\pm0.02$ & $-4.5\pm0.3$ & $0.01\pm0.01$ & $-4.8\pm0.5$ & $0.03\pm0.02$ & $-4.5\pm0.3$ & $0.02\pm0.02$ & $-4.8\pm0.5$&$<0.02$\tablenotemark{d}&$<-4.8$\tablenotemark{d}\\
0.15--0.20 & $0.03\pm0.02$ & $-4.4\pm0.2$ & $0.007\pm0.007$ & $-5.1\pm0.5$ & $0.02\pm0.01$ & $-4.8\pm0.3$ & $<0.01$\tablenotemark{d}& $<-5.1$\tablenotemark{d}&$<0.01$\tablenotemark{d}&$<-5.1$\tablenotemark{d}\\
0.20--0.25 & $0.10\pm0.02$ & $-3.96\pm0.09$ & $0.03\pm0.1$ & $-4.5\pm0.2$ & $0.04\pm0.02$ & $-4.5\pm0.2$& $0.02\pm0.01$& $-4.8\pm0.3$& $<0.007$\tablenotemark{d}& $<-5.3$\tablenotemark{d}\\
0.25--0.28 & $0.10\pm0.03$ & $-4.1\pm0.1$ & $0.016\pm0.011$ & $-4.9\pm0.3$ & $0.08\pm0.03$ & $-4.3\pm0.2$ & $0.02\pm0.02$ & $-4.9\pm0.5$& $0.01\pm0.01$& $-5.2\pm0.5$\\
0.28--0.31 & $0.17\pm0.02$ & $-3.32\pm0.05$ & $0.08\pm0.02$ & $-3.65\pm0.07$ & $0.14\pm0.02$ & $-3.49\pm0.06$& $0.05\pm0.01$ & $-4.0\pm0.1$ & $0.022\pm0.008$& $-4.4\pm0.2$\\
0.31--0.33 & $0.19\pm0.03$ & $-3.42\pm0.06$ & $0.08\pm0.02$ & $-3.78\pm0.09$ & $0.16\pm0.03$ & $-3.56\pm0.07$ & $0.06\pm0.02$ & $-4.1\pm0.1$& $0.015\pm0.009$& $-4.7\pm0.3$\\
0.33--0.35 & $0.19\pm0.03$ & $-3.58\pm0.07$ & $0.08\pm0.02$ & $-4.0\pm0.1$ & $0.14\pm0.03$ & $-3.78\pm0.09$ & $0.10\pm0.03$ & $-4.0\pm0.1$& $0.02\pm0.01$& $-4.7\pm0.3$\\
0.35--0.37 & $0.18\pm0.05$ & $-4.0\pm0.1$ & $0.05\pm0.02$ & $-4.5\pm0.2$ & $0.22\pm0.06$ & $-3.9\pm0.1$ & $0.10\pm0.04$ & $-4.3\pm0.2$& $<0.012$\tablenotemark{d}&$<-5.25$\tablenotemark{d}\\
0.37--0.41 & $0.16\pm0.03$ & $-3.94\pm0.07$ & $0.08\pm0.02$ & $-4.26\pm0.09$ & $0.25\pm0.04$ & $-3.76\pm0.05$ & $0.18\pm0.03$ & $-3.98\pm0.07$& $0.02\pm0.01$& $-4.9\pm0.2$\\
0.41--0.44 & $0.13\pm0.02$ & $-3.91\pm0.07$ & $0.04\pm0.01$ & $-4.4\pm0.1$ & $0.21\pm0.03$ & $-3.73\pm0.06$ & $0.15\pm0.03$ & $-3.92\pm0.07$ & $0.014\pm0.007$& $-4.9\pm0.2$\\
0.44--0.47 & \nodata\tablenotemark{e}& \nodata\tablenotemark{e} & \nodata\tablenotemark{e} & \nodata\tablenotemark{e} & $0.23\pm0.04$ & $-3.94\pm0.07$ & $0.20\pm0.04$ & $-4.04\pm0.08$ & $0.04\pm0.02$& $-4.7\pm0.2$\\
0.47--0.49 & \nodata\tablenotemark{e} & \nodata\tablenotemark{e} & \nodata\tablenotemark{e} & \nodata\tablenotemark{e} & \nodata\tablenotemark{e} & \nodata\tablenotemark{e} & $0.11\pm0.05$ & $-4.6\pm0.2$ & $0.07\pm0.04$& $-4.8\pm0.2$\\
0.49--0.52 & \nodata\tablenotemark{e} & \nodata\tablenotemark{e} &\nodata\tablenotemark{e} & \nodata\tablenotemark{e} & \nodata\tablenotemark{e} & \nodata\tablenotemark{e} & $0.19\pm0.04$ & $-4.18\pm0.08$ & $0.02\pm0.01$& $-5.1\pm0.2$\\
0.52--0.56 & \nodata\tablenotemark{e} & \nodata\tablenotemark{e} &\nodata\tablenotemark{e} & \nodata\tablenotemark{e} & \nodata\tablenotemark{e} & \nodata\tablenotemark{e}& $0.21\pm0.03$ & $-4.12\pm0.06$& $0.02\pm0.01$ & $-5.0\pm0.2$
\enddata
\tablecomments{
\tablenotetext{a}{Half-light radius, either along the major axis ($R_e$) or circularized ($R_{e,c}$), is in units of kpc.}
\tablenotetext{b}{Redshift intervals correspond to the redshift distribution of overdensities in F2 (Figure~\ref{f5}).}
\tablenotetext{c}{Number density $n$  in units of Mpc$^{-3}$.}
\tablenotetext{d}{For a given compactness definition there are no compact galaxies in this redshift interval.}
\tablenotetext{e}{For a given limiting stellar mass range the F2 galaxy sample is not complete in this redshift interval.}
}
\end{deluxetable*}

\subsection{ Summary of Global Evolutionary Constraints}\label{global}

Redshift evolution in the typical size, average  stellar population age, and  absolute and fractional abundances for galaxies segregated by  stellar mass and size (or the combination of the two) constrains the drivers of the size growth of the quiescent galaxy population. In Sections~\ref{smr},~\ref{dn}, and \ref{nde} we examine the evolutionary trends  that quiescent $M_\ast>10^{10}\, M_\sun$ galaxies follow for $0.1<z<0.6$. Here we summarize the constraints.

The size evolution of quiescent F2-HSC galaxies in three stellar mass bins (Figure~\ref{f10}B, Table~\ref{table3}) shows that the most massive galaxies ($M_\ast>10^{11}\, M_\sun$) grow much more slowly in size than the least massive ($M_\ast\sim10^{10}\, M_\sun$) galaxies. In stark contrast with the results at higher redshift \citep{vanderWel2014}, the size evolution of intermediate-redshift massive galaxies slows down and the size growth of low-mass systems accelerates over the same redshift (lookback time) interval. 

The distributions of sizes for younger and older quiescent F2-HSC galaxies (Figure~\ref{f12}) and the distributions of D$_n4000$  (indicator of the average stellar population ages) for large and small quiescent intermediate-redshift galaxies (Figure~\ref{f13}) provide additional clues about the mass-dependence of galaxy size growth. For $M_\ast>10^{11}\, M_\sun$ quiescent F2-HSC galaxies, the size distributions are insensitive to population age and the age distributions are insensitive to size.  In contrast, for $M_\ast<10^{11}\, M_\sun$, statistical tests demonstrate that the distributions of sizes for younger and older (and the distributions of galaxy ages for larger and smaller) lower-mass galaxies cannot originate from the same underlying distribution. The age distribution of high-mass quiescent galaxies regardless of their size suggests simple passive evolution from $z\sim0.6$ to $z\sim0.1$. On the other hand, the population of quiescent galaxies with lower stellar masses show age distributions dependent on galaxy size: on average, smaller galaxies are up to 2~Gyr older than their larger counterparts of the same mass. These mass-dependent trends are consistent with the downsizing scenario where the most massive galaxies complete their evolution at earlier times.

The evolution in number density and fraction within the general quiescent population  is a function of stellar mass  (Figures~\ref{f14}~and~\ref{f15}).  Massive galaxies with $M_\ast>10^{11}\, M_\sun$) have a constant absolute number density and a monotonically declining fractional abundance with decreasing redshift. Quiescent systems with lower stellar masses display  mass- and size-dependent evolution in both absolute and fractional abundance. Larger ($R_e>1.8$~kpc) galaxies with $M_\ast<10^{11}\, M_\sun$ increase their absolute number density as the redshift decreases but the fractional abundance remains approximately constant or increases (for $M_\ast\sim10^{10}\, M_\sun$ quiescent systems). Over the same redshift interval, smaller quiescent systems in the same mass range show constant absolute number density but they are a steadily decreasing fraction of the population. At smaller sizes and stellar masses, only the $R_e\sim1$~kpc, $M_\ast\sim10^{10}\, M_\sun$ quiescent systems show approximate constancy in both absolute and fractional abundance.

With some dependence on definition, different samples of massive compact quiescent galaxies in the F2-HSC sample (Figure~\ref{f16}, Table~\ref{table4}) the absolute and fractional number densities evolve in parallel with the most massive objects in the F2-HSC survey. The fraction of massive compact systems increases from $z\sim0.1$ to $z\sim0.5$. The number density of dense quiescent systems closely follows the distribution of overdensities in the F2 field without no significant monotonic increase towards the redshift limit of the survey, $z \sim 0.6$.

\section{Discussion}\label{dis}

The observed evolutionary trends constrain the effect that different mechanisms have on the observed size growth of the quiescent galaxy population at intermediate redshift (Section~\ref{global}). The next step is to relate the changes in galaxy size, age, number density and fractional abundance with redshift and their dependence on stellar mass and size of a galaxy to the proposed growth mechanisms. These mechanisms include both processes that affect individual quiescent systems and global trends that propel the evolution of the quiescent population as an ensemble (Section~\ref{gm}). In Section~\ref{lim} we consider the limitations of the redshift survey approach. Because the distribution of structure in the field is imprinted in all redshift trends we observe, we first explore the effect of cosmic variance on the comparisons we make with other surveys (Section~\ref{cv}). 

\subsection{Cosmic Variance}\label{cv} 

Redshift distributions of number densities/fractional abundances for the quiescent galaxy population are influenced by the distribution of structures in the field regardless of galaxy stellar mass or size (Figures~\ref{f14}~and~\ref{f15}). The rate of growth in average size for the most massive quiescent systems ($M_\ast>10^{11}\, M_\sun$) changes direction from (mildly) increasing to decreasing with redshift at the redshift of a large cluster complex (Figure~\ref{f11}B). The abundance of massive ($M_\ast>3\times10^{10}\, M_\sun$) compact quiescent galaxies depends heavily on the distribution of structure in the field (Figure~\ref{f16}). Thus it is important to investigate the impact of cosmic variance on the comparison between galaxy counts and (consequently) average size growth rates for quiescent galaxies in different surveys.

Similarity between  the quiescent galaxy number density in SHELS F2 survey and in the low-resolution spectroscopic survey PRIMUS, covering an area of 5.5~deg$^2$ distributed over five separate fields, confirms that the number density measurements in the 4~deg$^2$ F2 field are not significantly biased by cosmic variance (Section~\ref{masslim}). Furthermore, single median size measurements for $z<0.5$ 3D-HST quiescent systems (a 0.25~deg$^2$ field) with $10^{10}\, M_\sun<M_\ast<10^{11}\, M_\sun$ follow the trends in median size with redshift that we trace by sampling intermediate-redshift quiescent F2 galaxies in narrow redshift bins (Section~\ref{smr}). 

The only significant difference between 3D-HST and F2-HSC estimates (at the highest stellar masses, $M_\ast>10^{11}\, M_\sun$) is fully accounted for by the difference in volumes probed by two surveys (Section~\ref{smr}). If the sizes of the most massive galaxies at $z\sim0.25$ are traced using similarly high-resolution imaging over a larger area \citep[e.g., COSMOS-DASH, based on HST~WFC3~H$_{160}$ imaging of $\sim0.5$~deg$^2$ area within COSMOS field, ][]{Mowla2018}, the median values and the size growth rate agree within the quoted uncertainties with the values for  $M_\ast>10^{11}\, M_\sun$ F2 quiescent systems  (Table~\ref{table3}). This comparison demonstrates that trends in size growth of quiescent F2 galaxies segregated by stellar mass (Figure~\ref{f10}) are robust to the details of the large-scale structure in these different surveys.

In contrast with the F2-HSC results, a recent study of intermediate-redshift galaxies based on $ugri$-band photometric survey of 150~deg$^2$ \citep[KiDS, ][]{deJong2015} suggests that $M_\ast\sim2\times10^{11}\, M_\sun$ quiescent systems experience faster size growth than their $M_\ast<10^{11}\, M_\sun$ counterparts over the $0<z<0.5$ redshift range \citep{Roy2018}. Although the absolute growth rate for  $M_\ast>10^{11}\, M_\sun$ KiDS galaxies is within $\pm 1\sigma$ of the F2-HSC and  COSMOS-DASH results, the difference in the growth rate of  quiescent systems as a function of stellar mass is opposite to the behaviour in the F2-HSC survey where galaxies of lower stellar mass experience more rapid size growth than their more massive counterparts. 

In comparison with F2-HSC, the sizes of KiDS galaxies are based on lower-resolution imaging in the $r-$band ($\sim0\farcs7\pm0\farcs1$). In contrast to F2 galaxies with spectroscopic redshifts, KiDS galaxy redshift estimates are based on a machine-learning technique and aperture photometry in four bands of the visible wavelength range. Their spheroid-dominated (early-type, quiescent) galaxy selection criteria include a) the S\'ersic index of the best-fit 2D surface brightness model ($n>2.5$), and b) the SED fitting classification based on a set of 66 spectrophotometric \texttt{Le Phare} templates \citep{Ilbert2006}. 

Spectroscopy-based selection of quiescent galaxies in F2-HSC (Section~\ref{spec}) and photometric selection of quiescent systems in KiDS  introduce different levels of contamination by star-forming systems in the two samples, especially at stellar masses $\log(M_\ast/M_\sun)\lesssim10.75$ where spectroscopic information becomes more important \citep{Moresco2013}. Furthermore, a fraction of quiescent systems at $z\sim0$ and in higher redshift regimes have a (minor) disk component \citep[][and references therein]{Buitrago2018}. Differences between the characteristics of the two surveys and between the criteria used to extract quiescent samples likely contribute to the differences between trends in size growth of F2-HSC and KiDS systems.

\subsection{Growth Mechanisms}\label{gm}

\begin{figure*}
\begin{centering}
%\hspace*{-0.35in}
\includegraphics[scale=0.5]{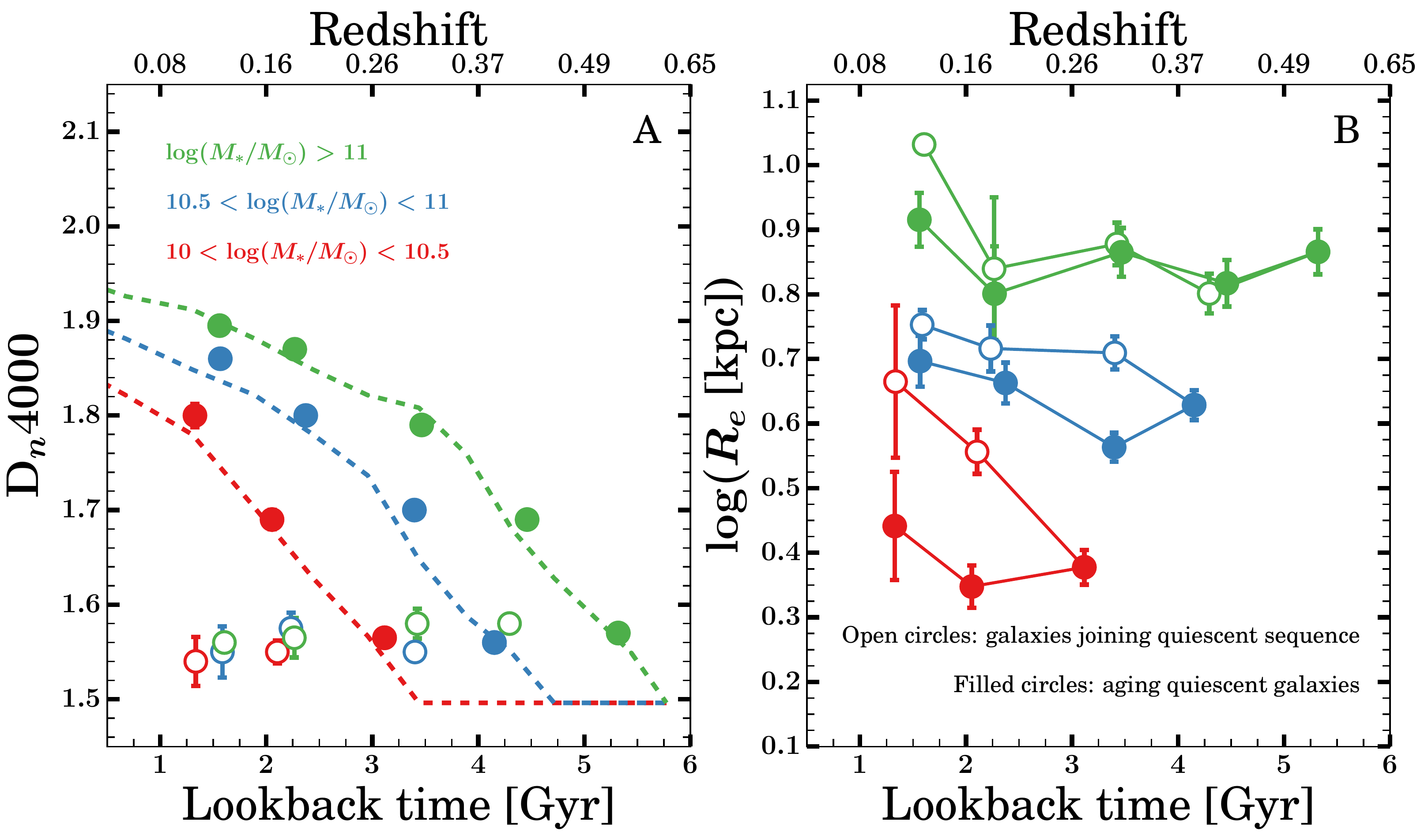}
\caption{A: D$_n4000$ index versus lookback time (redshift) for: 1) F2-HSC galaxies that become quiescent (i.e., have D$_n4000\sim1.5$) at the limiting redshift for each mass bin and age over the observed redshift interval (i.e., increase D$_n4000$  in steps that correspond to 1~Gyr time intervals based on the relation between D$_n4000$ and lookback time for model galaxy in Figure~\ref{f13}; filled circles), and 2) young quiescent F2-HSC systems  (with $1.5\leqslant D_n4000\leqslant1.6$) that  are coming onto the red sequence in each redshift interval and mass bin (open circles). Dashed colored lines show the redshift evolution of the D$_n4000$ index for a model galaxy (Figure~\ref{f13}) that became quiescent at the upper redshift limit for each of the three mass bins in panel. B: Size growth as a function of lookback time (redshift) for aging galaxies (filled circles) and young quiescent systems with $1.5\leqslant D_n4000\leqslant1.6$ (open circles). In both panels the three colors correspond to the stellar mass intervals introduced in Figure~\ref{f10}. \label{f17}}
\end{centering}
\end{figure*}

A comparison of HST-based surface brightness profiles of high-redshift ($z\sim2$) quiescent spectroscopic samples with the properties of radial profiles for similarly massive quiescent systems from the local volume \citep[SDSS samples; e.g.,][]{Belli2017} reveals a dramatic increase in the average size (normalized by stellar mass) for quiescent galaxies observed over large lookback times. The zero point of the stellar mass -- size relation  in the F2-HSC quiescent galaxy sample changes in agreement with the average growth rate derived over a broader redshift baseline \citep[][Figure~\ref{f11}A]{vanderWel2014}. The growth rates of intermediate-redshift quiescent systems of different stellar masses, however, contrast sharply with the  mass-dependent growth rates at higher redshift: at $z<0.6$ size growth slows down for the most massive galaxies ($M_\ast>10^{11}\, M_\sun$) and accelerates for less massive systems ($10^{10}\, M_\sun<M_\ast<3\times10^{10}\, M_\sun$, Figure~\ref{f10}).

Two classes of evolutionary processes  may contribute to the increase in size of quiescent galaxies: a) growth of individual galaxies that are already in the quiescent population \citep[e.g.,][]{vanDokkum2010}, and b) the addition of more extended newly quenched galaxies to the quiescent population at later times \citep[progenitor bias; e.g,][]{Franx2008,Carollo2013}. Individual quiescent systems may grow through major mergers \citep[e.g.,][]{Kaviraj2014}, late accretion or minor mergers \citep{Naab2009,Oser2012,Hilz2012,Newman2012}, or by feedback-driven adiabatic expansion \citep{Fan2008,Fan2010}. The evolutionary trends in number density, fractional abundance, and the distribution of average stellar population ages for quiescent systems segregated by stellar mass and size at $0.1<z<0.6$ provide clues to the evolutionary processes that dominate the size growth.

The most massive ($M_\ast>10^{11}\, M_\sun$) galaxies  show no significant differences among the size distributions segregated by age and redshift. Some fraction of massive systems must experience individual size growth because the average size of the massive population grows slowly. However, the average age distribution of these massive F2-HSC galaxies is consistent with passive evolution over the redshift interval covered by the F2-HSC survey. A very few massive quiescent galaxies are becoming quiescent at later times. Together with their constant number density and increasing population fraction with redshift, the combined sizes and ages of the most massive galaxies are consistent with the well-known downsizing picture \citep[i.e.,  more massive galaxies quench and become quiescent at earlier epochs,][]{Cowie1996}. 

In contrast to the most massive galaxies, the intermediate-redshift evolution of quiescent systems builds in complexity at lower stellar masses. For $M_\ast<10^{11}\, M_\sun$ quiescent F2-HSC systems there is  a clear offset between the size distributions of older and younger objects: older quiescent galaxies are on average smaller than their younger similarly massive counterparts. Furthermore, the number density of the smallest (most compact) systems of all stellar masses remains approximately constant over the intermediate redshift range, with exception of narrow redshift intervals dominated by dense structures where the number densities  increase sharply. In contrast, fractional abundance of compact systems  increases steadily with redshift. This set of $M_\ast<10^{11}\, M_\sun$ galaxy properties corresponds to the expected outcomes of a combination of two global trends: 1) downsizing, or the build up of stellar mass function below the characteristic mass $M^*$ at $z<1$ \citep[where $M^*\gtrsim7\times10^{10}\, M_\sun$, e.g.,][]{Ilbert2013,Muzzin2013} and  2) the increase in size of newly quenched galaxies at later epochs (i.e., progenitor bias).

The relative contribution to quiescent galaxy size growth of stochastic processes  affecting individual galaxies, and global trends  altering average size of a galaxy population through addition of more extended objects to the quiescent population at later times may depend on stellar mass.  We thus divide the quiescent sample into three stellar mass bins (as in Figure~\ref{f10} and Table~\ref{table3}) and follow the change in median size along two different D$_n4000$ (i.e., average stellar population age) tracks. We note that these two subsets are only a fraction of the full quiescent sample. Thus the growth rates we obtain for these subsets are not identical to the rates we quote in Figure~\ref{f10} and Table~\ref{table3} for the complete quiescent sample segregated by stellar mass. Figure~\ref{f17}A illustrates the change in D$_n4000$ with lookback time/redshift along two evolutionary tracks, and Figure~\ref{f17}B shows  trends in the median size with lookback time/redshift for all three stellar mass intervals.   

Filled circles in Figure~\ref{f17}A show the redshift evolution of D$_n4000$ index for a population that is passively evolving. For this experiment  the highest redshift (maximum lookback time) where we can observe each of the mass bins is the adopted start of the period of quiescence. We select quiescent systems with $D_n4000\sim1.5$ at this maximum redshift. As the redshift (lookback time) decreases, we select subsets of quiescent systems with D$_n4000$ corresponding to the relation between D$_n4000$ and redshift (lookback time) for the model galaxy in Figure~\ref{f13}. Figure~\ref{f17}A shows the match between the selected data (filled circles) and the passive evolution model (dotted line).  Figure~\ref{f17}B shows the median measured size of $D_n4000$ selected objects as a function of lookback time (solid dots connected by solid lines). The maximum change in the size of quiescent galaxies on this evolutionary track, corresponding to individual size growth, is $\sim0.15$~dex. If we attribute individual size growth to mergers, these processes increase the size of intermediate-mass systems in the sample up to $\sim 40\%$ over 2~Gyr of lookback time (blue filled circles in Figure~\ref{f17}B). At the high-mass end, the processes behind individual size growth produce an increase in size of $\sim20\%$ over 4~Gyr of lookback time (green filled circles in Figure~\ref{f17}B).

The change in the  median size of quiescent galaxies that join the quiescent population over this redshift interval (open circles in Figures~\ref{f17}A~and~\ref{f17}B) results from downsizing and progenitor bias. We track this evolutionary path by identifying objects with $1.5\leqslant D_n4000\leqslant1.6$ at each redshift (lookback time). Open circles in Figure~\ref{f17}A shows the average D$_n4000$ as a function of redshift for this selection. In  Figure~\ref{f17}B we trace the average size of this selected population with open circles.  The entry of recently quenched galaxies onto the quiescent population  has the largest effect in the lowest mass sample (red open circles in Figures~\ref{f17}A~and~\ref{f17}B), with median size increasing by almost a factor of 2. On the other end of the stellar mass distribution, galaxies with $M_\ast>10^{11}\, M_\sun$ (open green circles in Figures~\ref{f17}A~and~\ref{f17}B) entering the quiescent population show very little (if any) change in median size (note that the median size at a lookback time of ~1 Gyr is based on only one massive object).

Figure~\ref{f17} shows that galaxies with $M_\ast\sim10^{10}\, M_\sun$ experience the most prominent  size evolution at $0.1 \lesssim z \lesssim 0.35$. The evolution of these galaxies is driven by global processes that change the stellar mass and size distributions of the galaxies that enter the quiescent population at these redshifts. Their size growth rate is further enhanced  by processes like mergers, which increase the size of individual low-mass systems. In contrast, massive ($M_\ast\sim10^{11}\, M_\sun$) systems have mostly completed their size growth by $z\sim0.6$; the evolution in their fractional abundance is thus a manifestation of downsizing. These trends are broadly consistent with the properties of the local quiescent galaxy population \citep{Zahid2017}.

\subsection{Limitations}\label{lim}

Using the F2-HSC dataset we explore the  significance of 1) the growth of individual quiescent systems, and 2) the size increase (and mass decrease) of star forming galaxies with time  for the average size evolution of quiescent population as a function of stellar mass and size. Isolating the discrete individual contributions of different processes that drive size growth of individual quiescent systems requires additional measurements and deeper spectro-photometric surveys. 

The addition of velocity dispersion measurements provide an additional probe to constrain the evolution of quiescent galaxies \citep{Hopkins2010a,Zahid2017}. Discrimination between the relative importance of mergers with different mass ratios requires deep imaging, ideally accompanied by complete spectroscopic surveys to minimize observational uncertainties. However, redshift evolution in the number of galaxy pairs from these surveys must be converted into the evolution in merger rate \citep[e.g., ][]{Mundy2017}. At this step theoretical predictions of merger timescales based on a range of orbital parameters and galaxy dynamical states introduces additional uncertainties \citep[that can reach an order of magnitude,][]{Hopkins2010b}. 

Trends in size with redshift at $z\lesssim0.5$ show a mass dependence that is in stark contrast with the mass-dependent trends at $z>1$ (Section~\ref{smr}, Figures~\ref{f10}~and~\ref{f11}B). The growth rates of mass-segregated quiescent galaxies apparently change in the redshift range $0.5<z<1$: size growth slows down for $M_\ast>10^{11}\, M_\sun$ systems and simultaneously accelerates at for $M_* \sim 10^{10}$ galaxies. The limiting magnitude, $R=20.6$, limits our investigation to  $z\lesssim 0.5$. Deeper spectroscopic surveys could further trace and test these changes in the rate of  size growth for quiescent galaxies with different stellar masses at $z\lesssim1$. 

Throughout the redshift range of the survey, the HSC $i-$band images we use to measure galaxy sizes correspond to a range of rest-frame wavelengths, from $\sim5000$~\AA\ at $z\sim0.6$ to $\sim8000$~\AA\ at $z\sim0.1$. The expected color gradients in galaxies over this wavelength interval might induce additional redshift evolution in galaxy size. To estimate the amplitude of the color gradient effect and its possible impact on the results, we apply the empirical formula from \citet{vanderWel2014} to correct all galaxy sizes to their value at $5000$~\AA\ in the rest frame. The median corrected-to-uncorrected size measurement ratio for the F2-HSC sample is $1.033$. The correction is a function of redshift only; it does not depend on the D$_n4000$ index. Using size measurements with this (small) correction for the color gradient effect does not change any of the results.

Gradients in the stellar population within individual quiescent galaxies might also introduce systematics. The Hectospec aperture diameter (of $1\farcs5$) allows us to probe the average stellar population age within physical radial distances of $\sim3.5$~kpc from the galaxy center at the median redshift of F2-HSC galaxies $z\sim0.31$. Radial stellar population age gradient would be most prominent in objects with D$_n4000\sim1.5$ that are just joining quiescent sequence. Comparison between D$_n4000$ measurements for $z\lesssim0.25$ SHELS F2 galaxies that have  both SDSS-based (i.e., $3\arcsec$ fiber aperture) and Hectospec-based spectra shows that the two D$_n4000$ values are essentially identical \citep{Fabricant2008}. The absence of any variation in D$_n4000$ measurements between the two apertures suggests that the underlying stellar population (in both star-forming and quiescent systems) does not vary significantly over a factor of two in radial scales. These comparisons suggest that unless galaxies have much steeper population gradients than those in the SDSS comparison sample, the results are insensitive to this issue.

Figures~\ref{f14}~and~\ref{f15} display clear trends in the number density and relative fraction of quiescent systems correlated with overdensities in the field. The redshift evolution in the absolute and fractional abundance of massive compacts closely follows the large scale structure in the region (Figure~\ref{f16}). Because the richest galaxy clusters in F2 lie at $z\sim0.3$, the number densities of massive compacts peak at this redshift, decreasing by up to an order of magnitude at later and earlier epochs. To reduce the impact of cosmic variance a similar spectroscopic survey over a much larger area is required. For example, HectoMAP is a dense redshift survey of a $\sim53$~deg$^2$ field with median redshift $z=0.39$ \citep{Geller2016}. Follow-up HSC imaging, currently available for $\sim7$~deg$^2$ of the HectoMAP footprint \citep{Sohn2017,Sohn2018}, will cover the entire region \citep{Aihara2017}. Structural analysis of quiescent HectoMAP galaxies based on high-quality HSC images will be minimally affected by field-to-field variations as the impact of cosmic variance on this survey is reduced to $\sim4\%$ \citep{Driver2010}.

On its own, the F2-HSC survey confirms that massive compact quiescent galaxies are most abundant in the densest environments \citep{Damjanov2015b,FerreMateu2017,Buitrago2018}. However, based on stellar mass, size, and D$_n4000$ indices, we cannot distinguish between the processes that create and destroy these extreme systems.  In some cases gravitational lensing could reveal the extent of their dark matter halos \citep[e.g.,][]{Monna2015,Monna2017} and thus constrain the impact of tidal stripping on the formation of massive compacts in overdense regions. Deeper imaging surveys will also provide observational limits on the effects of minor mergers/accretions in growing (and thus destroying) massive quiescent compact systems in clusters.

\section{Conclusions}\label{conc}

By combining high-resolution HSC imaging with the magnitude-limited spectroscopic survey of a 4~deg$^2$ field \citep[SHELS F2,][]{Geller2014}, we explore the contribution of different types of evolutionary processes to the size growth of massive quiescent galaxies ($M_\ast>10^{10}\, M_\sun$) at intermediate redshift ($0.1<z<0.6$). We measure structural parameters, including size, by modelling surface brightness profiles of SHELS F2 galaxies in $i-$band HSC images with single S\'ersic profiles. We combine these measurements with spectroscopic properties, redshift and D$_n4000$ index (a proxy for average age of galaxy stellar population age), and estimates of stellar mass based on SED fitting, to examine the relations between galaxy size, stellar mass, and age and their evolution.

The parent sample of $\sim11000$ SHELS F2 galaxies displays a clear trend in average size with galaxy age. Divided into cells defined by a combination of narrow redshift slices encompassing similar volumes and bins of stellar mass ($\Delta[\log(M_\ast/M_\sun)]=0.5$~dex), star forming galaxies have on average larger sizes than their quiescent counterparts throughout the mass range $9.5<\log(M_\ast/M_\sun)\lesssim11.9$ and the redshift interval. To explore the change in size as galaxies evolve through  quiescence, we use $\sim3500$ galaxies with $M_\ast>10^{10}\, M_\sun$ and D$_n4000>1.5$ to trace:
\begin{itemize}
\item the intermediate redshift evolution  of the relation between stellar mass and size of quiescent systems as a function of their stellar mass;
\item the relation between galaxy size and average stellar population age at fixed mass and its redshift evolution;
\item redshift trends in galaxy age or size distributions for quiescent systems of similar stellar mass divided further into subsamples at median size or age, respectively;
\item the evolution in absolute and fractional abundance of quiescent galaxies segregated by stellar mass and size since $z\sim0.5$;
\item the size growth of quiescent galaxies that either a) become quiescent (i.e., reach D$_n4000=1.5$) at $z \sim 0.6$ and evolve further (i.e. increase D$_n4000$) towards $z\sim0.1$, or b) move onto the quiescent population (i.e., have D$_n4000\sim1.5$) throughout the survey redshift interval.
\end{itemize}

The relative contributions of different processes that drive quiescent galaxy evolution depend critically on galaxy stellar mass. Intermediate-redshift evolutionary trends for $M_\ast>10^{11}\, M_\sun$ galaxies differ from the trends with redshift for galaxies of lower stellar mass:

\begin{itemize}
\item The average size of  $M_\ast>10^{11}\, M_\sun$ quiescent systems increases by $25\%\, (\pm10\%)$ from $z\sim0.6$ to $z\sim0.1$ (green circles and green shaded area in Figure~\ref{f10}). At  $M_\ast<10^{11}\, M_\sun$ the evolution in average size accelerates with decreasing stellar mass; the size of $M_\ast\sim10^{10}\, M_\sun$ increases by more than 70\% from $z\sim0.3$ to $z\sim0.1$ (red circles and red shaded area in Figure~\ref{f10}).

\item For $M_\ast>10^{11}\, M_\sun$ galaxies, the size is insensitive to age at fixed stellar mass. For galaxies with $M_\ast<10^{11}\, M_\sun$ the size is strongly anti-correlated with age for a given stellar mass throughout the redshift range.

\item $M_\ast>10^{11}\, M_\sun$ quiescent galaxies the age distributions is insensitive to size and the size distribution is insensitive to age. In contrast, at lower stellar masses the age distribution for larger galaxies is offset towards younger ages relative to the age distribution for smaller galaxies. 

\item The number density of  $M_\ast>10^{11}\, M_\sun$ quiescent systems follows the redshift distribution of overdensities in the field and does not evolve with redshift. Their fractional abundance increases with redshift. At  $M_\ast<10^{11}\, M_\sun$ the evolution in number density and fractional abundance (within the general quiescent population at a given redshift) depend on both galaxy stellar mass and size. The number density of the largest galaxies decreases with redshift. This trend changes into constancy with redshift at intermediate sizes and then into an increasing trend with redshift for smallest galaxies. Galaxy sizes where these trend transitions occur decrease with decreasing stellar mass. The fractional abundance of lower mass systems follows similar stellar mass- and size-dependent trends. However, the abundances of lower mass quiescent systems continue to follow the redshift distribution of structures in the field.

\item There is very little growth ($\lesssim20\%$) in the average size of $M_\ast>10^{11}\, M_\sun$ galaxies that become quiescent at $z\sim0.6$ and age to $z\sim0.1$ (filled green circles in two panels of Figure~\ref{f17}) and the average size of similarly massive systems that are continuously joining the quiescent population over the same redshift interval (open green circles in two panels of Figure~\ref{f17}). This growth is consistent (within $1 \sigma$) with the expected change in average size of the total most massive quiescent population (green circles and green shaded area in Figure~\ref{f10}). At $M_\ast<10^{11}\, M_\sun$ the size growth of newly quenched quiescent galaxies depends on stellar mass and reaches $\sim80\%$ over 2~Gyr of cosmic time for the least massive $M_\ast\sim10^{10}\, M_\sun$ systems (open red circles in two panels of Figure~\ref{f17}). Quiescent systems with increasing D$_n4000$ display an average size growth that increases with stellar mass, reaching a maximum of $\sim35\%$ over 2~Gyr for $M_\ast\sim5\times10^{10}\, M_\sun$ galaxies (filled blue circles in two panels of Figure~\ref{f17}).
\end{itemize}

The evolutionary trends we observe depend critically on stellar mass and, at $M_\ast<10^{11}\, M_\sun$, depend partially on galaxy size. These dependences suggest that the processes dominating the intermediate-redshift evolution of quiescent systems are related to the changes in properties of the global galaxy population. Stochastic processes affecting individual quiescent systems play a secondary role that becomes more prominent at lower stellar masses:

\begin{itemize}

\item $M_\ast>10^{11}\, M_\sun$ quiescent galaxies have mostly completed their assembly by $z\sim0.6$. Both the addition of a small number of larger galaxies at later epochs (progenitor bias) and individual growth contribute equally to a mild increase in the average size (of $\lesssim25\%$). For the most massive F2 quiescent systems the only significant evolution is the increasing fractional abundance with redshift that is a consequence of downsizing.

\item At $M_\ast<10^{11}\, M_\sun$, individual size growth of quiescent systems through mergers becomes increasingly important for the quiescent size growth of galaxies with stellar masses $10.5<\log(M_\ast/M_\sun)<11$.

\end{itemize}

The mass-dependence of the average size growth rate we observe at $z<0.5$ contrasts directly with high-redshift observations. For galaxies at $z\gtrsim0.5$ in the 3D-HST+CANDELS survey \citep{vanderWel2014} size evolution is most rapid at large stellar masses ($M_\sun>10^{11}\, M_\sun$) and decelerates with decreasing mass. The difference between the trends in size growth rate with stellar mass at high and intermediate redshift highlights the importance of detailed spectro-photometric observation at $0.5<z\lesssim1$. In this transitional redshift interval, the relative contributions of the various  processes driving the average size evolution of quiescent systems may change significantly.

This study provides a baseline for future synergies between large-area high-resolution imaging campaigns (with e.g., HSC) and dense spectroscopic surveys (with e.g., Subaru Prime Focus Spectrograph). Spectro-photometric surveys of larger areas will further probe the relationships among the evolutionary trends we outline here and the distribution of large scale structure in the survey volume. Deeper spectroscopic campaigns will test the relative contribution of various processes to the growth of quiescent systems at lower stellar masses and at higher redshift. Deeper imaging surveys can provide statistical samples of low surface brightness companions of quiescent galaxies thus directly probing the importance of mergers with different mass ratios for the size growth of quiescent population at $z<1$. Targeted high-resolution imaging of strong gravitational lenses with known spectroscopic properties will probe the connection between the size growth of the luminous and dark matter content of quiescent systems in massive galaxy clusters. 

\acknowledgements

We thank the anonymous reviewer for providing thoughtful comments and suggestions that have improved the clarity of our manuscript.

We are very grateful to all of the Subaru Telescope staff and the HSC builders. We thank Dr. Okabe for sharing computer resources necessary for reduction and analysis of HSC images. I.D. acknowledges the support of the Canada Research Chair Program. H.J.Z. gratefully acknowledges the generous support of the Clay Postdoctoral Fellowship. M.J.G. and J.S. are supported by the Smithsonian Institution. We thank Saint Mary's University for supporting H.S. through the SMUworks undergraduate student employment program.

The creation and distribution of the NYU-VAGC is funded by the New York University Department of Physics. We are greatly indebted to the SDSS team as a whole, in the form of its builders, its participants, and its external participants, as well as the investment made by its sponsors. Funding for the creation and distribution of the SDSS Archive is provided by the Alfred P. Sloan Foundation, the Participating Institutions, the National Aeronautics and Space Administration, the National Science Foundation, the US Department of Energy, the Japanese Monbukagakusho, and the Max Planck Society. The SDSS Web site is at \url{http://www.sdss.org}. The SDSS is managed by the Astrophysical Research Consortium for the Participating Institutions. The Participating Institutions are the University of Chicago, Fermilab, the Institute for Advanced Study, the Japan Participation Group, the Johns Hopkins University, Los Alamos National Laboratory, the Max Planck Institute for Astronomy, the Max Planck Institute for Astrophysics, New Mexico State University, the University of Pittsburgh, Princeton University, the United States Naval Observatory, and the University of Washington.

{\it Facilities}: MMT (Hectospec) - MMT at Fred Lawrence Whipple Observatory and Subaru Telescope operated by the National Astronomical Observatory of Japan.

%\bibliography{F2sizes}

\end{document}